\documentclass[a4paper,11pt]{article}
\pdfoutput=1 

\usepackage{jcappub} 

\usepackage{graphicx}
\usepackage{dcolumn}
\usepackage{bm}
\usepackage{float}
\usepackage{amssymb}
\usepackage{amsmath}
\usepackage{epsfig}
\usepackage{url}
\usepackage{hyperref}
\usepackage{hhline}
\usepackage{color}
\usepackage{multirow}
\usepackage[normalem]{ulem}
\usepackage{slashed}

\usepackage[section]{placeins}
\definecolor{jd}{rgb}{0.858, 0.188, 0.478}

\def\lapp{\mathrel{\rlap{\raise.5ex\hbox{$<$}}
                    {\lower.5ex\hbox{$\sim$}}}} 
\def\gapp{\mathrel{\rlap{\raise.5ex\hbox{$>$}}
                    {\lower.5ex\hbox{$\sim$}}}}

{\newcommand{\lsim}{\mbox{\raisebox{-.6ex}{~$\stackrel{<}{\sim}$~}}}
{\newcommand{\gsim}{\mbox{\raisebox{-.6ex}{~$\stackrel{>}{\sim}$~}}}

\newcommand{\bmt}{\begin{pmatrix}}
\newcommand{\emt}{\end{pmatrix}}
\newcommand{\ba}{\begin{array}{c}}
\newcommand{\ea}{\end{array}}
\newcommand{\be}{\begin{equation}}
\newcommand{\ee}{\end{equation}}
\newcommand{\bea}{\begin{eqnarray}}
\newcommand{\eea}{\end{eqnarray}}

\newcommand{\bi}{\begin{itemize}}
\newcommand{\ei}{\end{itemize}}

\newcommand{\baz}{\begin{array}{cc}}

\newcommand{\mathsym}[1]{{}}

\newcommand{\bt}{\begin{tabular}}
\newcommand{\et}{\end{tabular}}

\newcommand{\benu}{\begin{enumerate}}
\newcommand{\eenu}{\end{enumerate}}

\newcommand{\bav}{\begin{array}{cccc}}

\affiliation[a]{Department of Physics, Indian Institute of Technology Guwahati, North Guwahati, Assam- 781039, India}
\affiliation[b]{State Key Laboratory of Theoretical Physics, Institute of Theoretical Physics, Chinese Academy of Sciences, Beijing 100190, P.R. China}

\author[a]{Basabendu Barman,}
\emailAdd{bb1988@iitg.ernet.in}
\author[a]{Subhaditya Bhattacharya,}
\emailAdd{subhab@iitg.ernet.in}
\author[b]{and Mohammadreza Zakeri}
\emailAdd{mzake001@ucr.edu}
\title{\bf Non-Abelian Vector Boson as FIMP Dark Matter} 

\abstract{In this analysis we demonstrate the freeze-in realization of a non-abelian vector boson dark matter (DM). We choose to elaborate an existing $SU(2)_N$ extension ($N$ stands for neutral) of the Standard Model (SM) with an additional $U(1)=S^{'}$ global symmetry, which stabilizes the vector boson ($X,\bar{X}$) as DM through unbroken $S=T_{3N}+S^{'}$ as the lightest odd $S$ particle. Apart from showing the right order of the $SU(2)_N$ coupling ($\sim 10^{-12}-10^{-13}$) required for the correct relic of DM via freeze in, the analysis reveals that the contribution to the freeze-in production of DM from the decay of a heavier scalar bi-doublet $\zeta_1^{0,\pm} \to \zeta_2^{0,\pm}X$ is equally important even after the decoupling of $\zeta_1^{0,\pm}$ from the thermal bath. This treatment of computing the relic abundance in context with freeze-in is practically model-independent and can be applied to all the scenarios where the DM is produced from the decay of a massive particle which was once in equilibrium with the thermal bath. This bi-doublet earlier was in equilibrium with the visible sector due to SM $SU(2)_L$ coupling. Moreover, the neutral component of $SU(2)_N$ scalar triplet ($\Delta$), responsible for neutrino mass generation in this framework, turns out to serve as additional DMs in the model and offers a multipartite freeze-in DM set up to explore. The allowed parameter space is obtained after estimating constraints from CMB, BBN and AMS-02 bound. This exercise nicely complements the freeze-out realization of ($X,\bar{X}$) as weakly interacting massive particle (WIMP) and distinguishes it through stable charge track signature at collider compared to leptonic signal excess as in WIMP scenario.
}

\begin{document}

\renewcommand*{\thefootnote}{\fnsymbol{footnote}}


\maketitle
\flushbottom


\setcounter{footnote}{0}
\renewcommand*{\thefootnote}{\arabic{footnote}}

\section{Introduction}
\label{sec:intro}
A particle dark matter (DM) is highly motivated from astrophysical observations. However, laboratory experiments like direct search and collider searches have not  detected any signals yet. It is therefore an important exercise to look for possibilities where DM interaction with the Standard Model (SM) is suppressed, yet produces correct relic density as has been dictated by anisotropies in Cosmic Microwave Background Radiation (CMBR). Among several possibilities~\cite{Hochberg:2014dra,Kuflik:2017iqs}, 
freeze-in mechanism serves as an interesting alternative~\cite{Hall:2009bx,Heeba:2018wtf,Elahi:2014fsa,Biswas:2018aib,Zakeri:2018hhe,Chakraborti:2019ohe}. 
In such a case, the DM is assumed not to be in equilibrium with thermal bath in the early universe owing to its tiny coupling to the visible sector. It is then produced then non-thermally via decay or annihilation of particles in thermal bath and freezes in once the temperature drops below DM mass
to yield correct relic density ($\Omega h^2\sim 0.1198$~\cite{Akrami:2018vks}). Several studies have been done in this direction to show that freeze-in can give rise to DM mass $\sim$ TeV scale, but the coupling with visible sector requires to be extremely tiny $\lsim 10^{-10}$. 
Therefore, such DM models are classified as feebly interacting massive particle (FIMP) (for a review see~\cite{Bernal:2017kxu}), as opposed to the thermal freeze-out of weakly interacting massive particle (WIMP).

Our aim here is to demonstrate the freeze-in of a non-abelian vector boson DM~\cite{KO201712} (see~\cite{Duch:2017khv} for an abelian example). We choose a well motivated $SU(2)_N$ extension ($N$ stands for neutral{\footnote{This means that the $SU(2)_N$ vector bosons are electromagnetic charge neutral.}}) 
of the Standard Model (SM) with an additional $U(1)=S^{'}$ global symmetry, which stabilizes the lightest of the vector boson ($X,\bar{X}$) as DM through unbroken 
$S=T_{3N}+S^{'}$ \cite{Fraser:2014yga}. Spontaneous symmetry breaking of $SU(2)_N$ results in massive gauge bosons. None of the SM particles are charged under $SU(2)_N$ 
and therefore $X,\bar{X}$ do not have a direct coupling to the visible sector besides  Higgs portal which is required to be tiny to avoid conflict with Higgs data. This naturally leads to the possibility of FIMP nature of $X,\bar{X}$ as DM. The model, in addition, possesses several features, for example, addressing neutrino mass generation via inverse seesaw of type-III irrespective of whether the DM is undergoing freeze-in or freeze-out and a possible unification to $SU(7)$.  

The DM $(X,\bar{X})$ in this model is produced from the decay of a scalar bi-doublet $\zeta_{1}^{0,\pm}$, owing to $SU(2)_N$ interaction\footnote{If the decay is kinematically forbidden, then annihilation of $\zeta_{1}^{0,\pm}$ to produce $(X,\bar{X})$ becomes important.}. 
Naturally, the correct relic density of $X,\bar X$ via freeze-in indicates that the $SU(2)_N$ gauge coupling needs to be $\sim\mathcal{O}(10^{-10})$. 
One of the salient features of this study is to show that the contribution to DM production from the decay of  $\zeta_{1}^{0,\pm}$ remains significant even after the freeze-out of  $\zeta_{1}^{0,\pm}$. We point out that such a feature is inevitable whenever the decay is slow enough, although not much elaborated in FIMP literature. Moreover, our formulae for computing the yield of the DM from the decay of a decoupled species is actually model-independent, and can be applied to any scenario where the freeze-in production of the DM is taking place via the decay of a heavy particle, which shared a common temperature with the thermal bath in the early universe. 
The subsequent change in the allowed parameter space due to the `late decay' turns out to be quite noteworthy.
The $SU(2)_N$ scalar triplet ($\Delta$) required for neutrino mass generation also provides with additional DM components in this model. 
The neutral components of $\Delta$ turn out to be stable at the scale of the universe life time, thanks to the small $SU(2)_N$ gauge coupling in the freeze-in mechanism advocated here. Therefore the model also serves as a multipartite FIMP DM set up, although the DM components do not have sizable interaction with each other.  
Hence, the freeze-in of each individual component remains unaffected by the presence of others, but provides with a sizable range of allowed parameter space to span the whole under abundant region in the absence of direct search constraint for these FIMP like DMs.

However, stringent bounds on the lifetime of semi-stable charged and neutral particles ($\zeta_{1,2}^{0,\pm}$) arises from Big Bang Nucleosynthesis (BBN), which essentially rules out hadronically decaying particle with lifetime $\tau>100~\rm sec$~\cite{Jedamzik:2006xz,Jedamzik:2009uy}. CMB, on the other hand, puts a lower bound on the lifetime of DM decaying to SM particles, which can potentially alter the ionization history (and hence the power spectrum) of CMB~\cite{Mambrini:2015sia}. Experiments like AMS-02~\cite{Aguilar:2016kjl} also puts lower bound on decaying DM from non-observation of anti proton excess. In our case, as we shall elaborate, bounds from CMB and AMS-02 are rather lose but BBN plays a crucial role by eliminating a large portion of the parameter space allowed from relic density condition.     

This same model from WIMP perspective, has already been thoroughly explored in~\cite{Fraser:2014yga,Barman:2018esi}. 
The present exercise therefore provides an opportunity to compare two different realizations (freeze out versus freeze in) of the same model. 
In the following, we see that this provides not only a distinction in terms of $SU(2)_N$ gauge coupling, but also in terms of DM mass. 
For example, in case of freeze-in, we are bound to stick to low DM mass: $m_X\lsim 50~\rm GeV$ 
(depending on the contribution of $X$ to total relic abundance), while the WIMP scenario is valid for DM mass even upto $\sim$ TeV. Finally, the distinction between the WIMP and FIMP realization can also arise 
in collider signature of the model. For WIMP case, it was shown that hadronically quiet single and two lepton channels could verify the existence of such a model framework. On the contrary, in FIMP realization, the signature can arise through stable charge tracks or displaced vertices of $\zeta_{1,2}^{\pm}$ as demonstrated here.

The paper is organised as follows: in Sec.~\ref{sec:model} we have provided the details of the model including the symmetry breaking and spectrum of the physical particles that are important for the present phenomenology. Then 
in Sec.~\ref{sec:numass} we have shown how light neutrino mass can be generated via inverse seesaw mechanism in this framework. Sec.~\ref{sec:darksect} contains the main DM analysis under which in subsection 
~\ref{sec:yields} and~\ref{sec:beq} we have discussed in detail the yield for $X$ and $\Delta$ by solving the Boltzmann equation (BEQ) and in subsection~\ref{sec:bbn} we have elaborated the impact of BBN and CMB bounds 
on the parameter space of this model. In Sec.~\ref{sec:collider} we have shown the possible signatures that this model may yield at the colliders. Finally in Sec.~\ref{sec:concl} we have summarized our findings.

%
%
%
%

\section{The Model}
\label{sec:model}

%



We have considered a $SU(2)_N$ extension of the SM ($N$ stands for neutral), where the lightest of the gauge bosons acts as a DM candidate. The particle content is chosen in such a way so that the spontaneous symmetry breaking (SSB) of $SU(2)_N$ to yields massive gauge bosons and at the same time it is also possible to generate correct light neutrino mass successfully as proposed in~\cite{Fraser:2014yga}. All the SM fermions are singlet under the new $SU(2)_N$. The stability of DM is ensured by an imposed global $U(1)$ symmetry ($S^{'}$), such that $S=S^{'}+T_{3N}$ remains unbroken.

\begin{table}[htb]
\begin{center}
\begin{tabular}{|c| c| c| c|c|c|} 
\hline
Particles & $SU(3)_c$ & $SU(2)_L$ & $U(1)_Y$ & $SU(2)_N$ & $S^{'}$ \\ [0.5ex] 
\hline\hline
$X_{1,2,3}$ & 1 & 1 & 0 & 3 & 0 \\
\hline
$n=\left(n_1,n_2\right)_{L,R}$ & 1 & 1 & 0 & 2 & 1/2 \\
\hline
$\chi\equiv\left(\chi_1,\chi_2\right)$  & 1 & 1 & 0 & 2 & 1/2 \\
\hline
$\zeta\equiv\begin{pmatrix}
\zeta_1^0 & \zeta_2^0\\
\zeta_1^- & \zeta_2^-
\end{pmatrix}$  & 1 & 2 & -1/2 & 2 & -1/2 \\
\hline
$\Delta\equiv\begin{pmatrix}
\Delta_2/\sqrt{2} & \Delta_3\\
\Delta_1 & -\Delta_2/\sqrt{2}
\end{pmatrix}$ & 1 & 1 & 0 & 3 & -1\\
\hline
$\Phi\equiv\begin{pmatrix}
\phi^+\\
\phi^0
\end{pmatrix}$ & 1 & 2 & 1/2 & 1 & 0\\
\hline\hline
\end{tabular}
\end{center}
\caption{Relevant particle content of the model and their charges under $SU(3)_c\otimes SU(2)_L\otimes U(1)_Y \otimes SU(2)_N \otimes S^{'}$. $SU(2)_L$ doublets are 
indicated by vertical parenthesis with $T_{3L}=\pm 1/2$ for up and down components respectively. $SU(2)_N$ doublet is depicted by entries in horizontal parenthesis with $T_{3N}=\pm 1/2$ to left and right fields respectively.}
\label{tab:charge}
\end{table}

The new particles introduced in the model and their charges under $SU(3)_C \,\otimes$ $SU(2)_L\, \otimes$ $U(1)_Y \otimes$ $SU(2)_N \otimes\, S^{'}$ are noted in Tab.~\ref{tab:charge}. 
In the gauge sector, there are three $SU(2)_N$ gauge bosons $X_{1,2,3}$, where $X(\overline{X}) = \frac{X_1 \mp iX_2}{\sqrt{2}}$ turns out to be mass degenerate and serves as DM component(s) 
of the model. In the fermion sector, three families of Dirac fermion doublets $n=\left(n_1,n_2\right)_{L,R}$ (under $SU(2)_N$) are introduced which mediate the interactions of the dark sector 
(non-zero $S$ charged particles as noted below) with the SM sector. The scalar sector consists of one $SU(2)_L$ scalar doublet $\Phi$ (which contains the 125 GeV Higgs boson), one $SU(2)_N$ doublet $\chi$, 
one scalar bi-doublet $\zeta$ and one $SU(2)_N$ scalar triplet $\Delta$. 
The Dirac fermion doublets, together with the scalar triplet ($\Delta$) are required for generating light neutrino masses, which shall be discussed in the next section. 
The scalar doublet and bi-doublets participate in spontaneous symmetry breaking (SSB) to generate masses for all the particles involved in the model. The minimization condition along with 
the physical states that appear after SSB are elaborated below. It is important to note here that $SU(2)_N\otimes S^{'}\to S(=S'+T_{3N})$ occurs via the non-zero vacuum expectation value (VEV) of $SU(2)_N$ scalar doublet: 
$\langle\chi_2\rangle=u_2$. In Tab.~\ref{tab:scghrg} we have tabulated the $S$ charge assignments for the new particles added in the model.
All the SM particles have zero $S$ charge. Therefore, particles with non-zero $S$ charge will be protected from decaying into the SM. 
We can assume $X (\bar{X})$ to be the lightest of the particles with non-zero $S$ charge to qualify as DM candidate(s). 
Furthermore, neutral components of scalar triplet $\Delta_{1,2,3}$ can be stable (if $SU(2)_N$ coupling is assumed to be very small for successful freeze-in of $X(\bar{X})$) and be part of a multi-component DM framework.

\begin{table}[htb]
\begin{center}
\begin{tabular}{|c| c| c| c|c|c|} 
\hline
Particles & $S^{'}$ & $S=S^{'}+T_{3N}$ \\ [0.5ex] 
\hline\hline
$X(\bar X)$ & 0 & +1(-1) \\
$X_3$ &  & 0 \\
\hline\hline
$n_{1L,R}$ & 1/2& +1 \\
$n_{2L,R}$ & & 0\\
\hline\hline
$\chi_1$ & 1/2 & +1\\
$\chi_2$ &  & 0\\
\hline\hline
$\zeta_1$&-1/2&-1\\
$\zeta_2$&&0\\
\hline\hline
$\Delta_1$&&-2\\
$\Delta_2$&-1&-1\\
$\Delta_3$&&0\\
\hline\hline
\end{tabular}
\end{center}
\caption{$S$ charge assignment for the new particles added in the model as introduced in Tab.~\ref{tab:charge}.}
\label{tab:scghrg}
\end{table}

The scalars which acquire VEV are: $\langle\chi_2\rangle=u_2$, $\langle \zeta_2^0\rangle =v_2$, $\langle\Delta_3\rangle =u_3$, and $\langle \phi^0\rangle=v_1$. 
Note that the VEV assignment is different here from the $SU(2)_N$ extension considered in~\cite{Bhattacharya:2011tr}, where $\langle \Delta_1^0 \rangle$ is also non-zero. One should  note here that this 
particular assignment of VEVs for different scalar multiplets of the model is the only possibility to keep the $SM\times S$ symmetry intact, so that VEVs are assigned to only those neutral scalars which have zero 
$S$-charge; exception of which can lead to breaking of the $S$-charge spoiling the underlying symmetry of the Lagrangian and render the vector boson DM unstable. For example, $SU(2)_N$ breaking is only possible through $\langle\chi_2\rangle$. 
This, in turn, ensures that only $\zeta_2$ can acquire a non-zero VEV due to the term $(\mu_1 \tilde{\Phi}^\dagger \zeta \chi + H.c.)$ in the scalar potential (see Eq.~\ref{eq:scalarpot}). Similarly, it is straightforward 
to show from the term $\left(\mu_2 \tilde{\chi}^\dagger \Delta \chi + H.c.\right)$, that only $\Delta_3$ can acquire a non-zero VEV to keep $S$-symmetry unbroken. One can also check by computing the 
Hessian{\footnote{The Hessian ($\mathcal{H}$) is a multi-dimensional matrix in the field space of the scalar potential, consisting of second-order partial derivatives with respect to the fields ($\phi_i$): 
$\mathcal{H}=\frac{\partial^2 V}{\partial \phi_i\partial\phi_j}$.}} of the scalar potential that with our choice of the VEVs, the potential satisfies the {\it necessary} condition to be at its local minimum.

With this VEV choices, $X_{1,2}$ bosons have equal masses in this model, and more importantly $S=S'+T_{3N}$ global symmetry remains unbroken unlike in~\cite{Bhattacharya:2011tr} as mentioned in the last paragraph. 
Also note that $U(1)_{EM}$ remains unbroken after the electroweak symmetry breaking as no scalar with EM charge receives a non-zero VEV. This ensures that the photon remains massless.
The masses of the other gauge bosons are then given by:

\begin{equation}
\begin{split}
m_W^2 = \frac{1}{2} g_2^2\left(v_1^2+v_2^2\right), \quad\quad
m_{X}^2 = \frac{1}{2}g_N^2\left(u_2^2+v_2^2+2u_3^2\right),  \quad\quad
m_{Z'}^2 \simeq  \frac{1}{2}g_N^2\left(u_2^2+v_2^2+4u_3^2\right),
\end{split}
\label{eq:xmass}
\end{equation}

where $Z-Z'$ mixing matrix is given by:

\begin{equation}
m_{Z,Z'}^2 = \frac{1}{2}\begin{pmatrix}
\left(g_1^2+g_2^2\right)\left(v_1^2+v_2^2\right) & -g_N\sqrt{g_1^2+g_2^2}\, v_2^2\\
-g_N\sqrt{g_1^2+g_2^2}\, v_2^2 & g_N^2\left(u_2^2 + v_2^2 + 4u_3^2\right)
\end{pmatrix}.
\end{equation}

The choice of the VEVs and couplings required in the present scenario is mainly dictated by the DM phenomenology. Since in this work we are interested to see the freeze-in aspect of the $SU(2)_N$ 
vector bosons (as mentioned in Sec.~\ref{sec:intro}), we require the gauge coupling $g_N$ to be extremely small ($\lsim\mathcal{O}(10^{-10})$) in order to keep the lightest gauge boson out of equilibrium. 
We will carefully evaluate the correct order of the coupling and explore effects of such small coupling in the DM phenomenology. An immediate consequence of this is that 
$g_N\lsim\mathcal{O}(10^{-10})$ results in a very large $u_2\sim\mathcal{O}(10^{10})~\rm GeV$ for $m_{X}\sim\mathcal{O}(\rm TeV)$. Therefore $v_2\ll u_2$ and $g_N\ll g_{1,2}$, 
which ensures small $Z$-$Z^{'}$ mixing~\cite{Andreev:2014fwa}. This hides $Z'$ of this model from being observed at the LHC, and adds to the freedom of 
choosing $m_{Z'}$ as a free parameter. This should again be contrasted to the case in~\cite{Barman:2017yzr}, where there is a minimum limit on $M_{X_{1,2,3}} \geqslant 1$ TeV, f
or the degenerate vector boson DM case to respect the bound from $Z^{'}$ search data. Furthermore, owing to $u_3\ll u_2$, it is clear that the $X$ boson masses are nearly degenerate, 
i.e. $m_{Z'}(m_{X_3})\simeq m_X$. \\

With this particle content at our disposal, we can write the most general scalar potential as \cite{Fraser:2014yga}:

\begin{equation}\label{eq:scalarpot}
\begin{split}
V &= \mu_\zeta^2 Tr(\zeta^\dagger \zeta) + \mu_\Phi^2 \Phi^\dagger \Phi + \mu_\chi^2 
\chi^\dagger \chi + \mu_\Delta^2 Tr(\Delta^\dagger \Delta) + (\mu_1 
\tilde{\Phi}^\dagger \zeta \chi + \mu_2 \tilde{\chi}^\dagger \Delta \chi + H.c.)\\ 
&+ {1 \over 2} \lambda_1 [Tr(\zeta^\dagger \zeta)]^2 + {1 \over 2} \lambda_2 (\Phi^\dagger \Phi)^2 + {1 \over 2} \lambda_3 Tr(\zeta^\dagger \zeta \zeta^\dagger \zeta) + {1 \over 2} \lambda_4 (\chi^\dagger \chi)^2 + 
{1 \over 2} \lambda_5 [Tr(\Delta^\dagger \Delta)]^2 \\ 
&+ {1 \over 4} \lambda_6 Tr(\Delta^\dagger \Delta - \Delta \Delta^\dagger)^2  + f_1 \chi^\dagger \tilde{\zeta}^\dagger \tilde{\zeta} \chi + f_2 \chi^\dagger \zeta^\dagger \zeta \chi + f_3 \Phi^\dagger \zeta \zeta^\dagger \Phi + f_4 \Phi^\dagger \tilde{\zeta} \tilde{\zeta}^\dagger \Phi   \\ 
&+ f_5 (\Phi^\dagger \Phi)(\chi^\dagger \chi) + f_6 (\chi^\dagger \chi) Tr(\Delta^\dagger \Delta) + f_7 \chi^\dagger 
(\Delta \Delta^\dagger - \Delta^\dagger \Delta) \chi + f_8 (\Phi^\dagger \Phi) Tr(\Delta^\dagger \Delta)\\ 
&+ f_9 Tr(\zeta^\dagger \zeta) Tr(\Delta^\dagger \Delta) + f_{10} Tr[\zeta(\Delta^\dagger \Delta - \Delta \Delta^\dagger) \zeta^\dagger],
\end{split}
\end{equation}
where
\begin{equation}
\tilde{\Phi}^\dagger = (\phi^0, -\phi^+), ~~~ \tilde{\chi}^\dagger = 
(\chi_2, -\chi_1), ~~~ \tilde{\zeta} = \begin{pmatrix}
\zeta_2^+ & - \zeta_1^+ \cr 
-\bar{\zeta}_2^0 & \bar{\zeta}_1^0\end{pmatrix}.
\end{equation}
Since $u_3, v_1, v_2 \ll u_2$ as argued above, the minimization conditions yield the following conditions for the VEVs to the scalar potential parameters:
\begin{align}
u_2^2  &\approx - \frac{\mu_{\chi}^2}{\lambda_4} \label{eq:sca:min}\\
v_1^2 &\approx -\frac{\mu_{\Phi}^2+f_5 u_2^2}{\lambda_2} \\
v_2 &\approx  -\frac{\mu_1 v_1 u_2}{\mu_{\zeta}^2 + f_1 u_2^2}\\
u_3 &\approx -\frac{\mu_2 u_2^2}{\mu_{\Delta}^2 + (f_6 + f_7) u_2^2} 
\end{align}
where from the last line it follows that unless $u_2 \ll \mu_{\Delta} $, we should have $\mu_2 \approx u_3$.\\

The SU$(2)_N$ triplet scalars $\Delta_{1,2,3}$, remain complex with masses given by\footnote{See Appendix~\ref{sec:scalar} for more details.}
\begin{align}
  m_{\Delta_1}^2 &\approx -4 f_7 u_2^2-\frac{2\mu _{2} u_2^2}{u_3},\\
  m_{\Delta_2}^2 &\approx -2 f_7 u_2^2-\frac{2 \mu_{2} u_2^2}{u_3},\\
  m_{\Delta_3}^2 &\approx \,\,  2\frac{u_2^2\, \mu _{2}}{u_3}.
  \end{align}
  
Note that since $\mu_{2} \approx u_3$, $\Delta_3$ is at the same scale as $u_2$, and $f_7$ determines the mass differences between $\Delta_1,~ \Delta_2~\rm{and}~\Delta_3$ as before~\cite{Barman:2018esi}. 
The rest of the neutral physical scalars of our model are: 
\begin{align}
  h &= \frac{1}{\sqrt{v_1^2 + v_2^2}} \left(v_1 \text{Re}(\phi_2) + v_2 \text{Re}(\zeta_2^{0})\, \right),\\
\xi_2^{0} &= \frac{1}{\sqrt{v_1^2 + v_2^2}} \left(-v_2\text{Re}(\phi_2) + v_1 \text{Re}(\zeta_2^{0})\, \right),\\
  \xi_1^0 &\approx \frac{1}{\sqrt{v_2^2 + u_2^2}} \left(-u_2 \zeta_1^0 + v_2 \chi_1 \right), \\
\eta^0 &\approx \frac{-1}{\sqrt{v_1^2 + u_2^2 ( 1 + v_1^2/v_2^2)}} \left(u_2 \text{Im}\left(\phi ^0_2\right) + \frac{u_2 v_1}{v_2} \text{Im}\left(\zeta ^0_2 \right) + v_1 \text{Im}\left(\chi _2 \right)\right),
\end{align}

with masses given by:
\begin{align}
m_{h}^2 &\approx  \,\,  4 \lambda_2 v_1^2,\\
m_{\xi_2^{0}}^2 &\approx  \,\, -\frac{2\mu _1 u_2 v_1}{v_2},\\
m_{\eta^0}^2 &\approx  -\frac{2 \mu _1 u_2 v_1}{v_2}\left(1+ \left(v_2/v_1\right)^2 \right),\\
m_{\xi_1^0}^2 &\approx 2 \left(f_2-f_1\right) \left(u_2^2+v_2^2\right) - \frac{2\mu _1 u_2 v_1}{v_2} -4 f_{10} u_3^2,\\
m_{\chi_2^R}^2 &\approx \,\,  4\lambda _4 u_2^2.
\end{align}

From the two charged scalars, $\zeta_1^-$ doesn't mix and is physical, while $H^+$ is an admixture of $\phi^+$ and $\zeta_2^+$:
\begin{align}
H^+ = \frac{1}{\sqrt{v_1^2 + v_2^2}} \left(v_2 \phi^+ + v_1 \zeta_2^+\right).
\end{align}
The masses of the charged scalars in terms of the scalar potential parameters and VEVs are as follows:
\begin{align}\label{eq:newscalarmass}
m_{\zeta_1^-}^2 &\approx  2(f_2  -f_1)u_2^2-\frac{2\mu_1 u_2 v_1}{v_2} -4 f_{10} u_3^2, \\
m_{H^+}^2 &= 2\left(v_1^2+v_2^2\right)\left(f_3-f_4-\frac{\mu _1 u_2}{v_1 v_2}\right).
\end{align}

The VEV $u_2$ being very large, dictates the scalar masses predominantly. Note that $\chi_2$ and $\Delta_{1,2,3}$ are at the same scale as $u_2$. For other scalars, dominant contribution arises from 
the presence of terms proportional to $\mu_1 u_2 v_1/v_2$ (except the SM Higgs). If we require the masses of these new scalars to be at TeV scale, we should have $m \sim 10 \sqrt{\mu_1 u_2} \sim 1$ TeV, 
which requires $|\mu_1| \lsim 10^{-9}$ GeV. In that case, $H^+, \eta^0, \xi_2^0$ can all be at $\mathcal{O}(1 \text{ TeV})$ scale. Without any further assumptions $\Delta_{1,2,3},~ \xi_1^0,~\zeta_1^-,~\chi_2^R$ 
are all around the $u_2$ scale and very heavy: $\mathcal{O}(10^{13} \text{ GeV})$. 

However, as we will demonstrate the freeze-in of $X$ is mainly dictated by $\zeta_1^{0,-}$ decays, and a large $m_{\zeta_1^{0,-}}$ mass will produce over abundance of the DM. In order to avoid the overabundance of $X$, 
we need $m_{\zeta_1^{0,-}} \lsim 10 \text{ TeV}$; which can be achieved by setting $f_2=f_1$ and cancelling the over powering $u_2$ term (see Eq. \ref{eq:newscalarmass}). 
In Tab.~\ref{tab:particles} we have listed the physical particles in our model (including SM) with their corresponding mass scales that fit the freeze-in requirement for the vector boson DM. It is 
also important to mention here that the correct DM relic density via freeze in of $X$ allows $m_X$ to vary in a large range between few GeV's to few hundred GeV's, but the heavier DM masses are constrained by BBN data. Therefore we have listed them in the range of $\mathcal{O}(1~\rm GeV)$ in Tab.~\ref{tab:particles}. Similarly, $n_{1,2}$ masses can also be as large as $\sim \mathcal{O}(10^5)$ GeV depending on the Yukawa coupling as we demonstrate in the next section although they have been classified to lie in $\mathcal{O}(1~\rm TeV)$ in the Tab.~\ref{tab:particles}.

\begin{table}[htb!]
\begin{center}
\scalebox{1.0}{
\begin{tabular}{|c|c|c|}
\hline
Scale & $|\mu_1| \lsim 10^{-9}$ GeV &$|\mu_1| \lsim 10^{-9}$ GeV, $f_1 = f_2$ \\
\hline\hline
$\mathcal{O}(10^{13}~\rm GeV)$ & $\Delta_{1,2,3},\zeta_1^-,\xi_1^{0},\chi_{2}^R$ & $\Delta_{1,2,3},\chi_{2}^R$ \\
\hline
$\mathcal{O}(1~\rm TeV)$ & $H^{+},\eta^0,\xi_2^0, n_1, n_2$ & $H^{+},\eta^0,\xi_1^{0},\xi_2^0,\zeta_1^-, n_1, n_2$\\
\hline
$\mathcal{O}(100~\rm GeV)$ & \multicolumn{2}{c|}{$W^{\pm}, Z, H, t$} \\
\hline
$\mathcal{O}(1~\rm GeV)$ & \multicolumn{2}{c|}{$X, Z', b, c, \tau$} \\
\hline
$\mathcal{O}(<1~\rm GeV)$ & \multicolumn{2}{c|}{$e, \mu, \nu_e, \nu_{\mu}, s, u, d$}\\
\hline
\end{tabular}
}
\end{center}
\caption{Spectrum of physical particles appearing at different scales assuming $g_N \sim 10^{-13}$, compatible with freeze-in requirement of $X(\bar{X})$. 
In the dark matter analysis we will further assume $f_1 \approx f_2$, such that both $\zeta_1^{-,0}$ are brought down to $\sim$ TeV scale.}
\label{tab:particles}
\end{table}


We would like to verify the alignment limit of the chosen vevs in the model. At low energies, we can identify SU(2)$_L$ scalar doublets $(\Phi, \tilde{\zeta_2})$ of our model with $(\Phi_1, \Phi_2)$ of the the usual two Higgs doublet model (2HDM)~\cite{Branco:2011iw,Bhattacharyya:2015nca,Haber:2013mia}. In our case $\tan\beta = v_2 / v_1 \ll 1$, while $\tan\alpha \approx -v_1/v_2$, where $\beta$ and $\alpha$ diagonalize the mass-squared matrices of the scalars and pseudoscalars respectively.  Therefore, we have an approximate alignment as in Type-I case of 2HDM i.e. $\cos(\beta - \alpha) \approx 0$, in the {\it decoupling limit} ($v_2 \ll v_1 \ll u_2$). We would like to emphasize here that this alignment results in recovering a CP-even scalar mass eigenstate with similar gauge, Yukawa and self interactions at tree level as those of the SM Higgs boson{\footnote{Note that, since no scalar induces a charge breaking VEV, hence the photon remains identically massless independent of the decoupling limit.}. The leading contribution to this approximation is given by:

\begin{equation}
\cos(\beta - \alpha) \approx (\sqrt{2} - 1) \frac{v_2}{v_1} + \mathcal{O} (\frac{\lambda_2 v_2^2}{\mu_1 u_2})  \leq 1.7 \times 10^{-3}
\end{equation}
which satisfies the CMS limits on type-I 2HDM safely lies within the CMS limits on Higgs couplings~\cite{Sirunyan:2018koj}. 

\section{Neutrino mass}
\label{sec:numass}

As already elaborated in~\cite{Fraser:2014yga,Barman:2018esi}, the generation of light neutrino mass is a novel feature of this model addressed together with DM. In principle, the light neutrino mass generation mechanism is independent of DM phenomenology. However, we will take a quick tour of the neutrino sector here and advocate a subtle 
phenomenological connection to the freeze-in prospect of the DM, in the light of mass scales introduced in Tab.~\ref{tab:particles}.

The gauge and $S$ invariant Yukawa terms responsible for neutrino mass generation in this model are given by:

\bea
f_{\zeta} &\left[ \left(\overline{\nu}_L \zeta_1^0 + \overline{e}_L \zeta_1^- \right) n_{1R} + \left( \overline{\nu}_L \zeta_2^0 + \overline{e}_L \zeta_2^- \right) n_{2R} \right] \label{f-zeta}\\
f_{\Delta} &\left[ n_1 n_1 \Delta_1 + \left(n_1 n_2 + n_2 n_1\right) \Delta_2/\sqrt{2} - n_2 n_2 \Delta_3  \right],\label{f-delta}
\label{eq:yukawaint}
\eea  

\begin{figure}[htb!]
$$
 \includegraphics[scale=0.36]{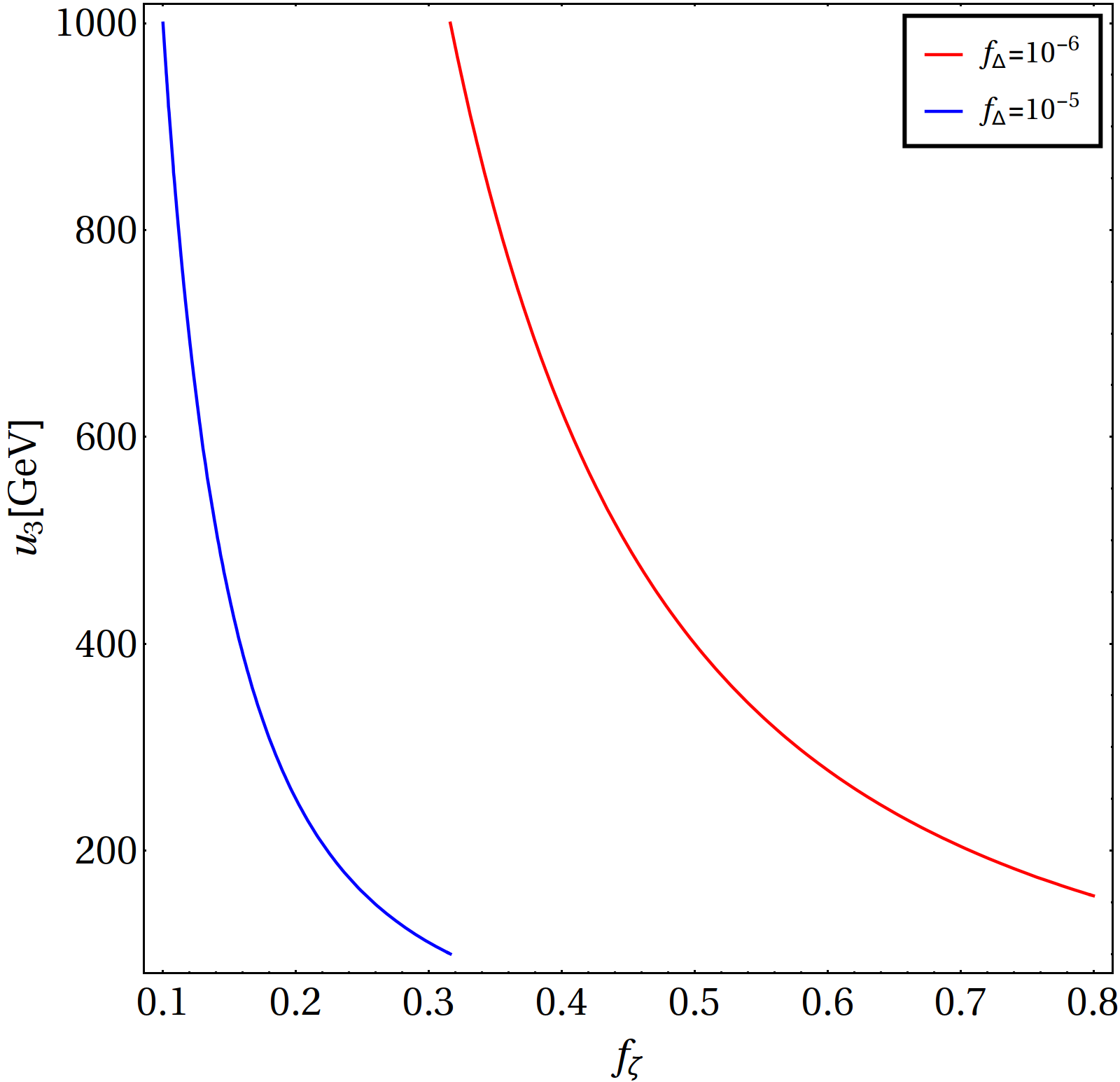}
  \includegraphics[scale=0.35]{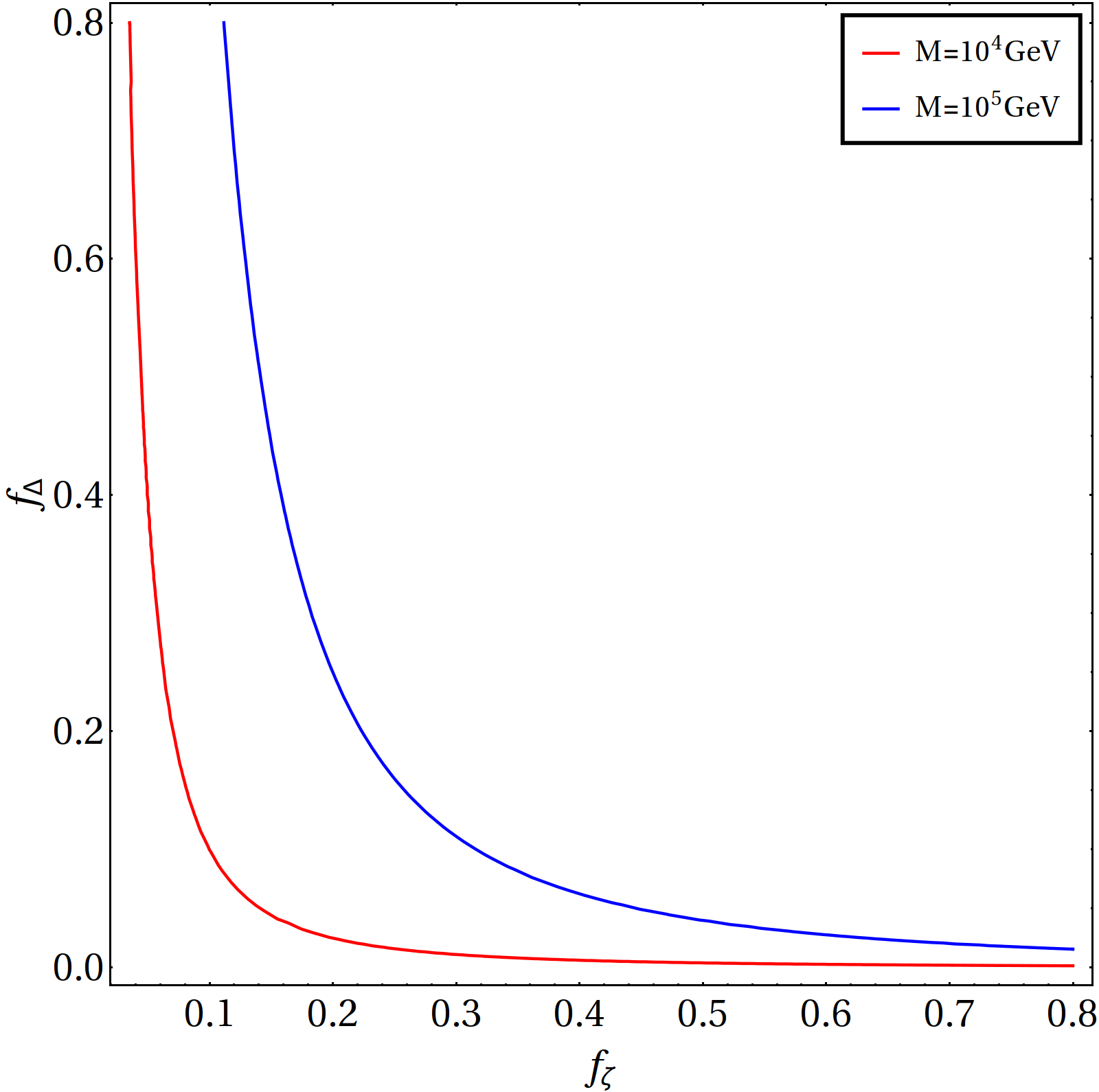}
$$
 \caption{LHS: Contours showing light neutrino mass $m_{\nu}\sim 0.1~\rm eV$ for different choices of the Yukawa coupling $f_{\Delta}=\{10^{-5},10^{-6}\}$ (in blue and red respectively) in $f_\zeta-u_3$ plane
 when $u_3\sim\mathcal{O}(100~\rm GeV)$ and $M\sim\mathcal{O}(\rm TeV)$. RHS: Contours satisfying light neutrino mass ($m_{\nu}\sim 0.1~\rm eV$) in $f_{\zeta}$-$f_{\Delta}$ plane for two different 
 choices of $n_{1,2}$ masses $M:\{10^4,10^5\}~\rm GeV$ (in red and blue curves respectively) with $u_3=100~\rm GeV$.}
\label{fig:numasscntr}
\end{figure}

The lepton number is conserved in~\eqref{f-zeta} with $n$ carrying $L=1$, and is broken to lepton parity, i.e. $(-1)^L$ by the $nn$ terms in~\eqref{f-delta}. After SSB we have the following mass terms for the neutrinos:

\begin{align}
f_{\zeta}\, v_2\, \overline{\nu}_L n_{2R} - f_{\Delta}^L\, u_3\, n_{2L} n_{2L} - f_{\Delta}^R\, u_3\, n_{2R} n_{2R}+ \text{h.c.}
\end{align} 
where $f_{\zeta}$ and $f_{\Delta}$ are $3\times 3$ matrices and $f_\Delta$ is further classified to address left handed ($f_\Delta^L$) and right handed ($f_\Delta^R$) Yukawa couplings separately. 
The neutrino mass matrix in the $\left(\overline{\nu}_L, n_{2R}, \overline{n}_{2L}\right)$ basis is then given by:

\begin{align}
M_{\nu} = \begin{pmatrix}
0 & m_D & 0\\
m_D & m_2' & M \\
0 & M & m_2
\end{pmatrix},
\end{align}

where each entry is a $3\times 3$ matrix with $m_D = f_{\zeta}\, v_2$, $m_2' = f_{\Delta}^R\, u_3$, $m_2 = {f_{\Delta}^{L}}^*\, u_3$, and $M$ is a free Dirac mass term in 
$M \left(\overline{n}_{2L} n_{2R} + \overline{n}_{2R} n_{2L}\right)$. Thus, the inverse seesaw neutrino mass is given in the form (assuming $f_{\Delta}^R\simeq f_{\Delta}^{L} \simeq f_\Delta$):

\bea
m_{\nu} \simeq \frac{m_D^2\, m_2}{M^2} = f_{\zeta}^2 f_{\Delta}\, \left(\frac{v_2}{M}\right)^2 u_3.
\label{eq:correctnumass}
\eea

From here we see that if we assume $u_3\sim\mathcal{O}\left(100~\rm GeV\right)$ and $M\sim\mathcal{O}\left(1~\rm TeV\right)$, we can generate light neutrino mass in the correct ballpark for $f_{\Delta}\sim\mathcal{O}\left(10^{-6}\right)$ and $f_{\zeta}\sim\mathcal{O}(1)$. This is shown in the LHS of Fig.~\ref{fig:numasscntr} where the two contours correspond to light neutrino 
mass $m_{\nu}\sim 0.1~\rm eV$ for smaller values of $f_{\Delta}:\{10^{-5},10^{-6}\}$ in blue and red respectively. One can however, choose a less fine-tuned $f_{\Delta}\sim\mathcal{O}\left(1\right)$ at the expense of making the RHNs super heavy $\sim 10^5$ GeV. This is depicted in the RHS of Fig.~\ref{fig:numasscntr} where we have chosen a fixed $u_3=100~\rm GeV$ and obtained contours of correct neutrino mass ($m_{\nu}\sim 0.1~\rm eV$) for two different choices of $M:\{10^4,10^5\}~\rm GeV$ (red and blue curves respectively) in the plane of $f_\zeta-f_\Delta$. We would like to mention here that the second choice of heavy $n_{1,2}$ is more desirable as it does not require the Yukawa couplings to be extremely fine-tuned, i.e. $\sim\mathcal{O}(10^{-6})$ and secondly will help us in addressing the freeze-in of $X(\bar{X})$ as the only possible decay mode of $\zeta_1^{\pm,0}$. More interestingly, it will also distinguish the collider signature of this model from that of the WIMP example~\cite{Barman:2018esi} (details in Sec.~\ref{sec:collider}).

\section{Dark Sector}
\label{sec:darksect}

In this set-up we assume $X$ to be the lightest non-zero $S$ charge particle and hence a DM candidate as stated in Sec.~\ref{sec:model}.
This choice is even more natural when $g_N$ is considered small as we have here for the freeze-in of $X$. Because of this, 
$X$ is not in thermal equilibrium in the early universe and is produced via the decay or annihilation of an odd-$S$ particle that can be in thermal bath. Therefore, in our model, 
$X$ can be produced via freeze-in from the decays of the scalar triplet $\Delta$ (Fig.~\ref{fig:loopdel3}) and the bi-doublet scalar 
 components $\zeta_1^{0,\pm}$ (Fig.~\ref{fig:zetaDecay}) when kinematically accessible. It is important to note that the decay occurs before and 
after the decoupling of the heavier particle from the thermal bath and we carefully illustrate how the `late decays' can contribute significantly to the relic density of DM. 
To be compatible with the required relic abundance via freeze-in, the coupling requires to be $g_N \sim 10^{-15} - 10^{-10}$, which in turn results in a very high mass scale for the $\Delta$'s as explained in Sec.~\ref{sec:model}. 
The mass hierarchy among different components of the $\Delta$ (namely, $\Delta_{1,2,3}$) is controlled by the parameter $f_7$. 
If we assume $f_7=0$ then all three components have the same mass. $\Delta$'s can be produced from the Higgs quartic interaction, which is a function of $f_8$. Therefore, freeze-in production of $\Delta$ requires $f_8\sim 10^{-12}$, while for freeze-out: $f_8\sim 1$. The Feynman graph for the production of $\Delta$ is shown in Fig.~\ref{fig:deltaprodc}. 
For simplification, we assume $f_{9,10} \ll f_8$ so that the $\Delta \leftrightarrow \zeta$ mixing and conversions can be neglected.  We will show that $\Delta$'s are naturally stable in the freeze-in scenario (for $g_N \sim 10^{-15} - 10^{-10},~f_8\sim 10^{-12}$), and we have $\{\Delta_{1,2,3}\}$ as long-lived relics that contribute to the DM abundance in addition to $X(\bar{X})$. Different production possibilities of $\Delta$, depending on the choice of the model parameters and possible degeneracy is classified in Tab.~\ref{tab:X:scen}.

\begin{table}[H]
$$
\begin{tabular}{c|c|c|c|c|}
\cline{2-5}
          & \multicolumn{2}{c|}{Degenerate $\Delta$ ($f_7 = 0$)} & \multicolumn{2}{c|}{Non-degenerate $\Delta$ ($f_7\neq 0$)} \\ 
          \hline

\multicolumn{1}{|c|}{Production mechanism of $\Delta$} & Freeze-Out  & Freeze-In & Freeze-Out     & Freeze-In    \\ \hline
\multicolumn{1}{|c|}{Scalar quartic coupling $f_8$} & $\mathcal{O}(1)$  & $\mathcal{O}(10^{-12})$ & $\mathcal{O}(1)$     & $\mathcal{O}(10^{-12})$    \\ \hline
\end{tabular}
$$
\caption{\label{tab:X:scen}Different production scenarios for $\Delta$'s, and their dependence on $f_{7,8}$ couplings. Throughout this study the freeze-in of $X$ is assumed.}
\end{table} 

We primarily focus on the degenerate scalar triplet scenario with $f_7=0$ and 
calculate the yield of the DM components ($\{X, \Delta_{1,2,3}\}$) via freeze-in by solving Boltzmann
equations (BEQ) in subsection.~\ref{sec:yields}.

\subsection{Degenerate $\Delta$'s with $f_8 \sim \mathcal{O}(10^{-12})$}
\label{sec:yields}

The masses of particles in $\Delta$ triplet are degenerate if $f_7=0$. In this case, $\Delta_{1,2,3}$ can decay via the tree level diagram on the LHS of Fig.~\ref{fig:loopdel3}. 
This diagram is only possible after the EWSB when $\Delta_3$ mixes with the Higgs via $f_8\left(\Phi^{\dagger}\Phi\right)tr\left(\Delta^{\dagger}\Delta\right)$. 
The effective vertex for {$\Delta_3\to t\bar{t}$ decay is $\sim f_8 m_t u_3/m_{\Delta_3}^2$. Since $m_{\Delta_3}\sim\mathcal{O}(10^{13}~\rm GeV)$ and 
$f_8\sim\mathcal{O}(10^{-12})$ in the freeze-in scenario, the decay vertex factor is tiny. This results in a very large lifetime for $\Delta_3$. As an estimate, 
if we set $m_{\Delta_3}=10^{13}~\rm GeV$, $f_8=10^{-12}$ and $u_3=100~\rm GeV$, this decay width turns out to be {$2\times 10^{-57}~\rm GeV$, which dubs into $\sim 6 \times 10^{32}~\rm sec$} in terms of decay lifetime. This is of course much larger than lifetime of the universe, which is $\sim 10^{17}~\rm sec$. This makes all of the $\Delta$'s stable 
\footnote{If we calculate the lifetime of $\Delta_1$ from the decay {$\Delta_1\to\bar{X}\bar{X}t\bar{t}$ turns out to be $6\times 10^{34}~\rm sec$} for similar choices of $g_N=10^{-13}$, $f_8=10^{-12}$ and $m_X=5~\rm GeV$.}.

\begin{figure}[htb!]
$$
\includegraphics[scale=0.35]{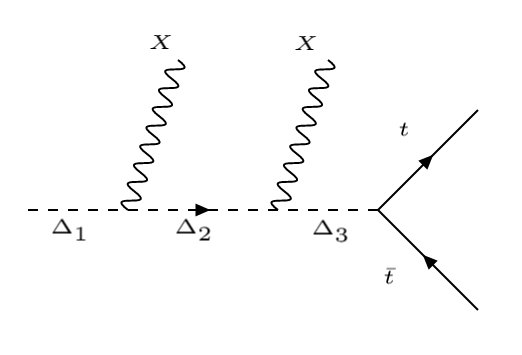}
\includegraphics[scale=0.35]{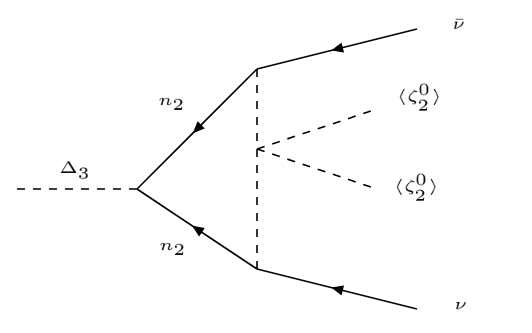}
$$
\caption{ Decay of $\Delta$'s at tree level to {$t\bar{t}$} (left) and decay of $\Delta_3$ via 1-loop to SM neutrinos (right).}
\label{fig:loopdel3}
\end{figure}

\begin{figure}[htb!]
$$
\includegraphics[scale=0.35]{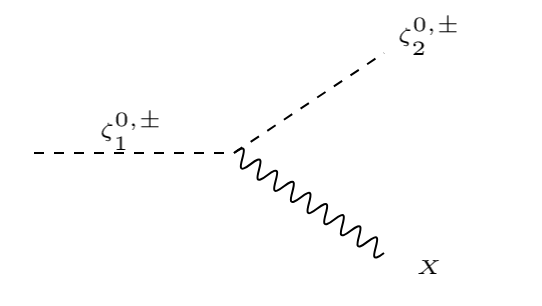}
\includegraphics[scale=0.35]{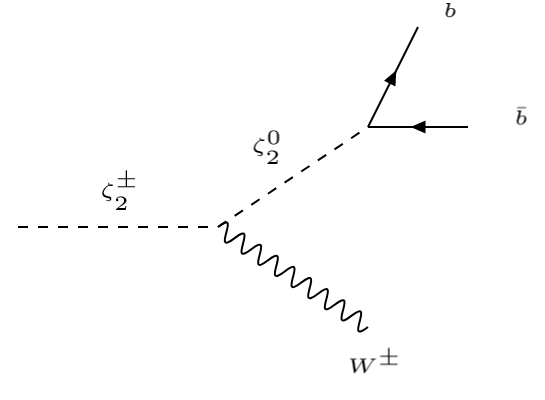}
$$
\caption{Left: Decay of $\zeta_1^{0,\pm}$ to $\zeta_2^{0,\pm}$ and $X$ resulting in production of $X$ via freeze-in for $g_N\sim\mathcal{O}\left(10^{-13}\right)$. Right: Decay of $\zeta_2^{\pm}$ to SM final states after $SU(2)_N$ is broken.}
\label{fig:zetaDecay}
\end{figure}

On the other hand, the loop-induced decay of $\Delta_3$ to SM neutrinos (RHS of Fig.~\ref{fig:loopdel3}), is possible even before the EWSB mediated by Yukawa couplings with $n_2$. 
The decay width for this process {\footnote{The amplitude for the loop diagram has been computed manually and cross-checked using {\tt Package-X}~\cite{Patel:2015tea}.}} is understandably 
small as it is proportional to the SM neutrino mass. Even if we assume the Yukawa couplings involved in this decay to be $f_{\zeta}\sim\mathcal{O}(1)$ and $f_{\Delta}\sim\mathcal{O}(10^{-6})$, 
the decay width turns out to be $\sim 4.75\times 10^{-65}~\rm GeV$ for $m_{n_2} \sim 1$ TeV and $m_{\zeta_2} \sim 200$ GeV.  Therefore, with a lifetime of $10^{41}~\rm sec$, we conclude that 
$\Delta_3$ (and hence $\Delta_{1,2}$) is always a long-lived relic. $X$ is kinematically stable and the prime DM candidate of the model with $m_X < m_{\zeta_1}, m_{\Delta_2}$. 
If we further assume $g_{N} \sim \mathcal{O}(10^{-13})$, then $\zeta_{1}^{0,-} \to \overline{X} + \zeta_2^{0,-}$ will produce $X$ via freeze-in mechanism (LHS of Fig.~\ref{fig:zetaDecay}). 
Note that after spontaneous breakdown of $SU(2)_N$, bi-doublet $\zeta$ breaks into two doublets, i.e. $\left(\zeta_1^0, \zeta_1^-\right)$ \&  $\left(\zeta_2^0, \zeta_2^-\right)$. 
After the electroweak SSB, $\zeta_2^{0,-}$ components will decay as shown in the RHS of Fig.~\ref{fig:zetaDecay}. As mentioned above, we are assuming $f_{9,10} \ll f_8$ 
for simplicity, such that $\Delta \leftrightarrow \zeta$ conversion is negligible. In the following subsection we calculate abundance for both $X$ and $\Delta$ using appropriate BEQ.

\subsubsection{Computation of yield for $\Delta$ and $X$}
\label{sec:beq}

Let us first estimate the yield of $\Delta$, which become stable due to small $f_8$ and contribute to DM relic density as elaborated in the last section. The rate of change of number density of $\Delta$ is governed by the following BEQ:

\begin{equation}
\begin{split}
\label{eq-C1-Bol-del-1}\dot{n}_{\Delta_i} + 3 H n_{\Delta_i} =& \int d\Pi_\Phi \, d\Pi_{\Phi^*} \, d\Pi_{\Delta_i} \, d\Pi_{\Delta_i^*} (2\pi)^4 \delta^4(p_{\Phi}+p_{\Phi^*} - p_{\Delta_i} - p_{\Delta_i^*})\\
\times& \left[ |\mathcal{M}|^2_{\Phi \Phi^*\to \Delta_i \Delta_i^*} \, f_{\Phi} f_{\Phi^*} (1+f_{\Delta_i})(1+f_{\Delta_i^*})\right.\\
&\, \left.- |\mathcal{M}|^2_{\Delta_i \Delta_i^* \to \Phi \Phi^*} \, f_{\Delta_i} f_{\Delta_i^*} (1+f_{\Phi})(1+f_{\Phi^*}) \right],
\end{split}
\end{equation}
where $d\Pi_i \equiv d^3p_i /(2\pi)^3 2E_i$ is the phase space factor, and number density is given by
\begin{equation}
n_i = \frac{g_i}{(2\pi)^3} \int d^3p\, f_i(p) ,
\end{equation}
where $g_i$ denotes the effective relativistic degrees of freedom and $H$ is the Hubble constant. Following Fig.~\ref{fig:deltaprodc}, 
the only way of producing $\Delta$ or depleting its number density occurs through the quartic interaction with Higgs ($H$). We are 
focusing on the freeze-in production of $\Delta$ before EWSB and after SSB of $SU(2)_N$. Higgs can then produce a pair of superheavy $\Delta$ only beyond the threshold center-of-mass (C.O.M) energy $s \ge 4m_{\Delta}^2$.


\begin{figure}
$$
\includegraphics[scale=0.45]{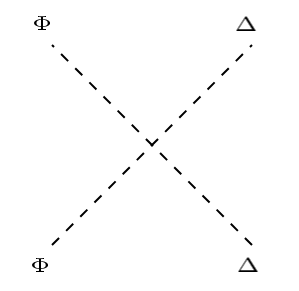} 
$$
\caption{Production of $\Delta$ via contact interaction from the annihilation of SM Higgs {\it before} EWSB. This channel is proportional to the coupling $f_8$.}
\label{fig:deltaprodc}
\end{figure}

Now, assuming negligible initial abundance for $\Delta_i$'s, we can set $f_{\Delta_i}=0$. We can also neglect the Pauli-blocking/stimulated emission effects, i.e. $f_i\ll 1$. Therefore, Eq.~\eqref{eq-C1-Bol-del-1} simplifies to
\begin{equation}
\begin{split}
\label{eq-C1-Bol-del-2}\dot{n}_{\Delta_i}  + 3 H n_{\Delta_i}  =& \frac{T}{512\, \pi^6} \int_{4m_{\Delta}^2}^{\infty} ds \, d\Omega \,|\mathcal{M}|^2_{\Phi \Phi^*\to \Delta_i \Delta_i^*} \, P_{\Phi\Phi^*}\,P_{\Delta_i\Delta_i^*} \, K_1(\sqrt{s}/T) /\sqrt{s},
\end{split}
\end{equation}
where, $T$ is the temperature, $s$ is the C.O.M energy of the production process $\Phi\Phi^*\to \Delta_i \Delta_i^*$ and we defined $P_{ij}$ as~\cite{Edsjo:1997bg}:
\begin{equation}
P_{ij} \equiv \frac{\left[s - (m_i+m_j)^2\right]^{1/2}\left[s - (m_i - m_j)^2\right]^{1/2}}{2\sqrt{s}} = \sqrt{\frac{s}{4} - m_i^2}.
\end{equation}
Since $|\mathcal{M}|^2_{\Phi \Phi^*\to \Delta_i \Delta_i^*} = f_8^2$, Eq.~\eqref{eq-C1-Bol-del-2} simplifies to:
\begin{equation}
\begin{split}
\label{eq-C1-Bol-del-3}\mathcal{S}\cdot \dot{Y}_{\Delta} =& \frac{T\cdot f_8^2}{512\, \pi^5} \int_{4m_{\Delta}^2}^{\infty} ds\,\sqrt{s-4m_{\Delta}^2}\sqrt{s-4m_{\Phi}^2}\, K_1(\sqrt{s}/T) /\sqrt{s},
\end{split}
\end{equation}
where the number density is converted to comoving yield, scaled as $Y_{\Delta} = n_{\Delta} / \mathcal{S}$ and $\mathcal{S}= 2\pi^2 g_{\star}^S\, T^3/ 45$ is the entropy 
and $m_{\Phi} = 0$. Using $\dot{T} \approx -T\cdot H$, and $x \equiv m_{\Delta}/T$ we rewrite Eq.~\eqref{eq-C1-Bol-del-3} as:
\begin{equation}
\begin{split}
\label{eq-C1-Bol-del-5}
\frac{dY_{\Delta}}{dx} =& \frac{45\, M_{\text{pl}}\,f_8^2\, x^3}{1024\,g_{\star}^S\sqrt{g_{\star}^{\rho}}\,1.66 \pi^7\, m_{\Delta}^4} \int_{4m_{\Delta}^2}^{\infty} ds\,\sqrt{s-4m_{\Delta}^2}\, K_1(x\cdot \sqrt{s}/m_{\Delta}),
\end{split}
\end{equation}
so the yield for $\Delta_i$ can be written as:
\begin{equation}
\begin{split}
\label{eq-C1-Y-del-1} Y_{\Delta_i}
=& \frac{0.10588\, M_{\text{pl}}\,f_8^2}{\pi^7\, m_{\Delta_i}} \int_0^{\infty} dx \, \frac{x^2}{g_{\star}^S\sqrt{g_{\star}^{\rho}}} \cdot K_1(x)^2 \approx 3.244 \times 10^{-7}  \left( \frac{M_{\text{pl}}}{m_{\Delta_i}} \right) \,f_8^2,
\end{split}
\end{equation}
where the last equality only holds if we assume a constant relativistic DOF $g_{\star} \approx 100$ during the freeze-in of $\Delta_i$'s. 
The total relic abundance of $\Delta$'s, i.e. $\Omega_{\Delta} = 3 \Omega_{\Delta_i}$ owing to its degeneracy is given by
\begin{equation}
\label{eq-C1-Omega-del-1} 
\Omega_{\Delta}\cdot h^2 = 3 \times \frac{2970 \, m_{\Delta}\, Y_{\Delta_i}^{\infty}\, \text{cm}^{-3}}{1.88\times 10^{-29}\, \text{g}\cdot\text{cm}^{-3}} \approx 10^{21} f_8^2
\end{equation}
We see that for $f_8 \in 6.19 \times [ 10^{-13}, 10^{-12}]$, we get relic density in the correct ball park, when the corresponding relic density of $\Delta$ is
$\Omega_{\Delta}\cdot h^2 \approx [\%1, \%100]$ of the observed DM relic density.


We now focus on $X$. The relic abundance of $X$ is coming from the $\zeta_1^{0,-}$ decays (see Fig. 3). We also assume that $n_{1,2}$ is heavier than 
$\zeta_1$, so that $\zeta_1 \to \zeta_2 +X$ constitutes 100\% decay branching fraction of $\zeta_1$.
The contributions from the $\Delta$ decays in Fig. 2, are exceedingly slow for small $f_8$. This makes $\Delta$ a DM candidate (as explained before) with negligible contribution to the freeze-in production of $X$. 
The production of $X$ from annihilation ($\zeta_1 \zeta_1 \to X X$) is much more suppressed due to the presence of $g_N^2$ in the amplitude and can be neglected. 
Therefore, the contribution to the number density of $X$ can then be written as:
\begin{equation}
\begin{split}
\label{eq-C1-Bol-X-1}\dot{n}_{\bar{X}} + 3 H n_{\bar{X}} =& 2\int d\Pi_{\bar{X}} \, d\Pi_{\zeta_1} \, d\Pi_{\zeta_2} (2\pi)^4 \delta^4(p_{\zeta_1} - p_{\zeta_2} - p_{\bar{X}})\\
 \times& \left[ |\mathcal{M}|^2_{\zeta_1\to \bar{X}+ \zeta_2} \, f_{\zeta_1} (1+f_{\bar{X}})(1+f_{\zeta_2}) - |\mathcal{M}|^2_{\zeta_2 + \bar{X} \to \zeta_1} \, f_{\zeta_2} f_{\bar{X}}(1+f_{\zeta_1}) \right]\\ 
 =& 2\int d\Pi_{\zeta_1} \, e^{-\left(E_{\zeta_1}-\mu_{\zeta_1}\right)/T'} \, e^{-(m_{\zeta_1}/E_{\zeta_1})\theta(t-t_D) \Gamma \cdot (t-t_D)} \left(2  m_{\zeta_1} g_{\zeta_1}\right) \, \Gamma_{\zeta_1\to \zeta_2 + \bar{X}},
\end{split}
\end{equation}
where $\theta$ is the step function, $t_D$ is the $\zeta$ decoupling time from the SM particles, and the factor of $2$ arises from same contribution of $\zeta_1^0$ and $\zeta_1^-$ decays to $X$ production. 
It is important to note that when $\zeta_1$ is in thermal equilibrium with the hot plasma, $T = T'$, but after $\zeta_1$ decouples: $T \neq T'$. 
We have set $f_{\bar{X}} = 0$ in accordance with freeze-in prescription and used:

\begin{equation}
\Gamma\equiv \Gamma_{\zeta_1\to \zeta_2 + \bar{X}} = \int \frac{1}{2 m_{\zeta_1}} \frac{|\mathcal{M}|^2_{\zeta_1\to \bar{X}+ \zeta_2}}{g_{\zeta_1}}  (2\pi)^4 \delta^4(p_{\zeta_1} - p_{\zeta_2} - p_{\bar{X}})\, d\Pi_{\bar{X}} \, d\Pi_{\zeta_2},
\end{equation}
where $g_{\zeta_1}=1$. Since $\zeta$ is a cold relic then $E_{\zeta_1} \approx m_{\zeta_1}$ for $t>t_D$, which simplifies Eq.~\eqref{eq-C1-Bol-X-1}:
\begin{equation}
\begin{split}
\label{eq-C1-Bol-X-2}\dot{n}_{\bar{X}} + 3 H n_{\bar{X}}
=& 2e^{-\theta(t-t_D) \Gamma\cdot(t-t_D)}  \int \frac{d^3p_{\zeta_1}}{(2\pi)^3\, \gamma_{\zeta_1}}\, e^{-\left(E_{\zeta_1}-\mu_{\zeta_1}\right)/T'} \, \Gamma_{\zeta_1\to \zeta_2 + \bar{X}}\\
=&  \frac{m_{\zeta_1}^2 \, \Gamma_{\zeta_1\to \zeta_2 + \bar{X}}}{\pi^2}\, T' \, K_1(m_{\zeta_1}/T')\, e^\frac{\mu_{\zeta_1}}{T'} e^{-\theta(t-t_D) \Gamma \cdot(t-t_D)} ,
\end{split}
\end{equation}
where $\gamma_{\zeta_1} = E_{\zeta_1} / m_{\zeta_1}$. Using the comoving yield $Y_{\bar{X}} = n_{\bar{X}} / \mathcal{S}$ and $\dot{T} \approx -T\cdot H$ we have:
\begin{equation}
\begin{split}
\label{eq-C1-Bol-X-4}\frac{dY_{\bar{X}}}{dT}=& -\frac{m_{\zeta_1}^2 \, \Gamma_{\zeta_1\to \zeta_2 + \bar{X}}}{\pi^2}\, \frac{T'}{T\, H\, \mathcal{S}} \, K_1(m_{\zeta_1}/T')\, e^\frac{\mu_{\zeta_1}}{T'} e^{-\theta(t-t_D) \Gamma \cdot(t-t_D)},
\end{split}
\end{equation}

The presence of $T^{'}$ in BEQ segregates the equation to the cases (i) before decoupling and (ii) after decoupling of $\zeta_1$ from thermal bath. 
It is important to relate temperature $T^{'}$ to the decoupling temperature ($T_D$). Throughout our calculations we use (see Appendix~\ref{sec:chempot}):
\begin{align}
 T' &= \left(\frac{R(t_D)}{R(t)}\right)^2 \cdot T_D, \qquad &m- \mu(T') =& \frac{T'}{T_D} \cdot  (m-\mu_D),
\end{align}
for cold relics. We now express $T'$ in terms of the temperature of the hot plasma ($T$) using the conservation of entropy, i.e. $S = \mathcal{S}\cdot R^3(t)=g_{\star}^S\, T^3\, R^3(t)=\text{const.}$\footnote{$g_{\star}^{S}$ and $g_{\star}^{\rho}$ are the effective number of relativistic degrees of freedom for entropy and energy respectively.}:
\begin{equation}
T \propto (g_{\star}^S)^{\frac{-1}{3}}\, R^{-1}.
\end{equation}
Now, since at the decoupling $T'_{D} = T_{D}$, we then have:
\begin{equation}
\label{eq:tempratio:cold} T' =  \left(\frac{g_{\star}^s(T)}{g_{\star}^s(T_D)}\right)^{\frac{1}{3}} \left(\frac{g_{\star}^{\rho}(T)}{g_{\star}^{\rho}(T_D)}\right)^{\frac{1}{4}} \left(\frac{T^2}{T_D}\right),
\end{equation}
Using the definitions of $\mathcal{S}$ and $H$ we get BEQ after decoupling as
\begin{equation}
\begin{split}
\label{eq-C1-Bol-X-5}\frac{dY_{\bar{X}}}{dT}\bigg|_{t>t_D}=& -\frac{90\, M_{\text{Pl}}\, m_{\zeta_1}^2 \, \Gamma}{1.66(4\pi^4)}\, \left(\frac{T'}{T^6}\right) \, \frac{K_1(m_{\zeta_1}/T')}{g_{\star}^S(T)\, \sqrt{g_{\star}^{\rho}(T)}}\, e^{\mu_{\zeta_1}/T'} e^{-\Gamma\cdot(t-t_D)}\\
=& -\frac{90\, M_{\text{Pl}}\, m_{\zeta_1}^2 \, \Gamma}{6.64 \pi^4}\, \left(\frac{1}{T^4\, T_D}\right) \, \frac{K_1(m_{\zeta_1}/T') \, e^{\mu_{\zeta_1}/T'}}{\left[g_{\star}^S(T_D) g_{\star}^S(T)^2 \right]^{1/3}\, \left[g_{\star}^{\rho}(T_D) g_{\star}^{\rho}(T) \right]^{1/4}} \\ &\times \exp \left[- \frac{0.301 \Gamma\, M_{\text{pl}}}{\sqrt{g_{\star}^{\rho}}} \left(\frac{1}{T^2} -\frac{1}{T_D^2} \right)\right],
\end{split}
\end{equation}
while for $t<t_D$ we have:
\begin{equation}
\begin{split}
\label{eq-C1-Bol-X-6}\frac{dY_{\bar{X}}}{dT}\bigg|_{t<t_D}=& -\frac{90\, M_{\text{Pl}}\, m_{\zeta_1}^2 \, \Gamma}{1.66(4\pi^4)\, T^5}\, \frac{K_1(m_{\zeta_1}/T)}{g_{\star}^S(T)\, \sqrt{g_{\star}^{\rho}(T)}}\, e^{\mu_{\zeta_1}/T},
\end{split}
\end{equation}
Changing the variable from $T \to x=m_{\zeta_1}/T$ yields:
\begin{align}
\frac{dY_{\bar{X}}}{dT} &= \frac{dx}{dT} \frac{dY_{\bar{X}}}{dx} = \frac{-x^2}{m_{\zeta_1}}\cdot \frac{dY_{\bar{X}}}{dx},\\
\frac{\mu_{\zeta_1}}{T'}&= \alpha(x, x_D) \left( \frac{x^2}{x_D}\right) - x_D,
\end{align}

where

\begin{equation}
\begin{split}
\alpha(x,x_D) \equiv &  \left[\frac{g_{\star}^s(x_D)}{g_{\star}^s(x)} \right]^{1/3}\, \left[\frac{g_{\star}^{\rho}(x_D)}{g_{\star}^{\rho}(x)} \right]^{1/4},\\
\eta(x,x_D) \equiv & \alpha(x,x_D)\, g_{\star}^s(x) \, \sqrt{g_{\star}^{\rho}(x)}.
\end{split}
\end{equation}

Using above, the BEQs before and after decoupling Eq.'s~\eqref{eq-C1-Bol-X-5} \& ~\eqref{eq-C1-Bol-X-6} become:

\begin{align}
\label{eq-C1-Bol-X-7}
\frac{dY_{\bar{X}}}{dx} =& \begin{cases}
\, \frac{90}{6.64 \pi^4\,g_{\star}^s(x) \, \sqrt{g_{\star}^{\rho}(x)}}\Sigma_{\zeta}\, x^3 K_1(x) , \qquad &x <  x_D \\ \\
\, \frac{90\, x^2 x_D\, \Sigma_{\zeta}}{6.64 \pi^4\, \eta(x,x_D)}K_1\left[\alpha(x,x_D)\, \left(\frac{x^2}{x_D}\right)\right]\, e^{ \frac{\left[\alpha(x, x_D)x^2-x_D^2\right]}{x_D}}e^{- \frac{0.301\, \Sigma_{\zeta} }{\sqrt{g_{\star}^{\rho}}}(x^2- x_D^2)}, \qquad  &x_D < x
\end{cases}
\end{align}
where we defined
\begin{equation}
\Sigma_{\zeta} \equiv \frac{M_{\text{Pl}}\, \Gamma}{m_{\zeta_1}^2}.
\end{equation}

\noindent The total $\bar{X}$ yield from $\zeta$ decays is given by:
\begin{equation}
\begin{split}
\label{eq-C1-Bol-X-8}Y_{\bar{X}}^{\infty}=& \frac{90\, \Sigma_{\zeta} }{6.64 \pi^4} \left( \int_0^{x_D} \frac{x^3 K_1(x)}{g_{\star}^s(x) \, \sqrt{g_{\star}^{\rho}(x)}}\,  dx \right.\\
&+\left. \int_{x_D}^{\infty}  e^{\frac{- 0.3\, \Sigma_{\zeta} }{\sqrt{g_{\star}^{\rho}(x)} } (x^2-x_D^2)}  \frac{x_D\cdot x^2}{\eta(x,x_D)} \, K_1\left[\alpha(x,x_D)\, \left(\frac{x^2}{x_D}\right)\right] e^{ \frac{\alpha(x, x_D)x^2-x_D^2}{x_D}}\, dx \right),
\end{split}
\end{equation}
The second integral within the parenthesis (from $x_D$ to $\infty$) indicates the contribution of the late decays of $\zeta_1$ (after its decoupling from thermal bath) to the production of DM ($X$). 
We would like to point out that an elaborate estimation of such contribution has not been carried out before in literature. The relic abundance of $X$ and $\bar{X}$ is then given by
\begin{equation}
\label{eq-C1-Omega-Tot} 
    \Omega_{\text{DM}}^X \cdot h^2= 2 \times\Omega_X \cdot h^2 = 2 \times\frac{2970 \, m_X\, Y_{X}^{\infty} \, \text{cm}^{-3}}{1.88\times 10^{-29}\, \text{g}\cdot\text{cm}^{-3}},
\end{equation}
where the factor of two is due to the contribution from both $X$ and $\bar{X}$. Note that, Eq.~\ref{eq-C1-Bol-X-8} is practically independent of the model concerned (except for the couplings and masses appearing in $\Sigma_{\zeta}$) 
as it solely involves the number of degrees of freedom $g_{*}^s$ and $g_{*}^\rho$ and the dimensionless quantity $x_D$ which for all practical purposes can be taken to be $x_D\sim 25$. Hence this is a generic expression for 
computing yield in all such cases where the DM is produced via freeze-in from the decay of a heavy species before and after its decoupling from the thermal bath.

\begin{figure}[htb!]
$$
\includegraphics[scale=0.7]{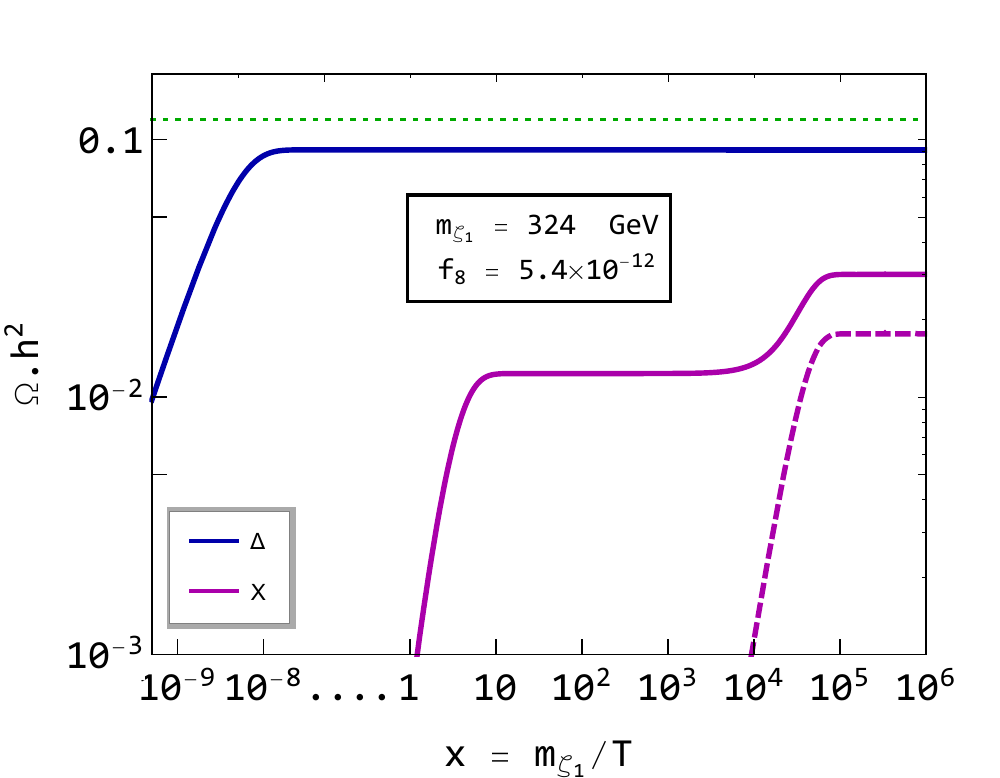}
\includegraphics[scale=0.7]{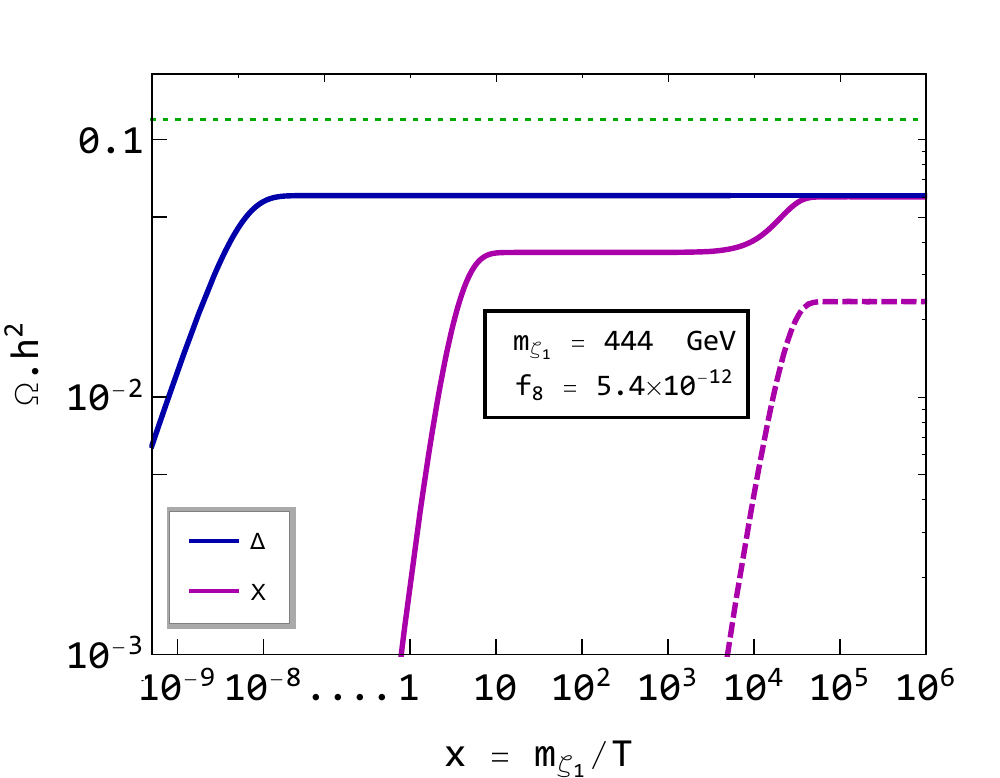}
$$
$$
\includegraphics[scale=0.7]{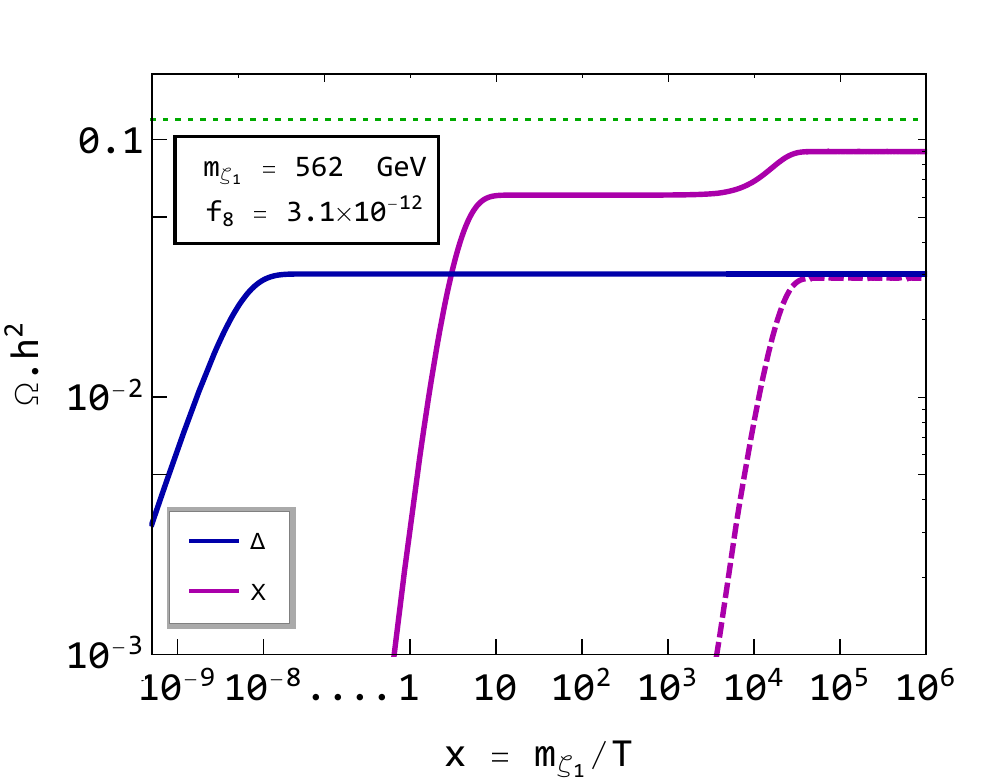}
\includegraphics[scale=0.7]{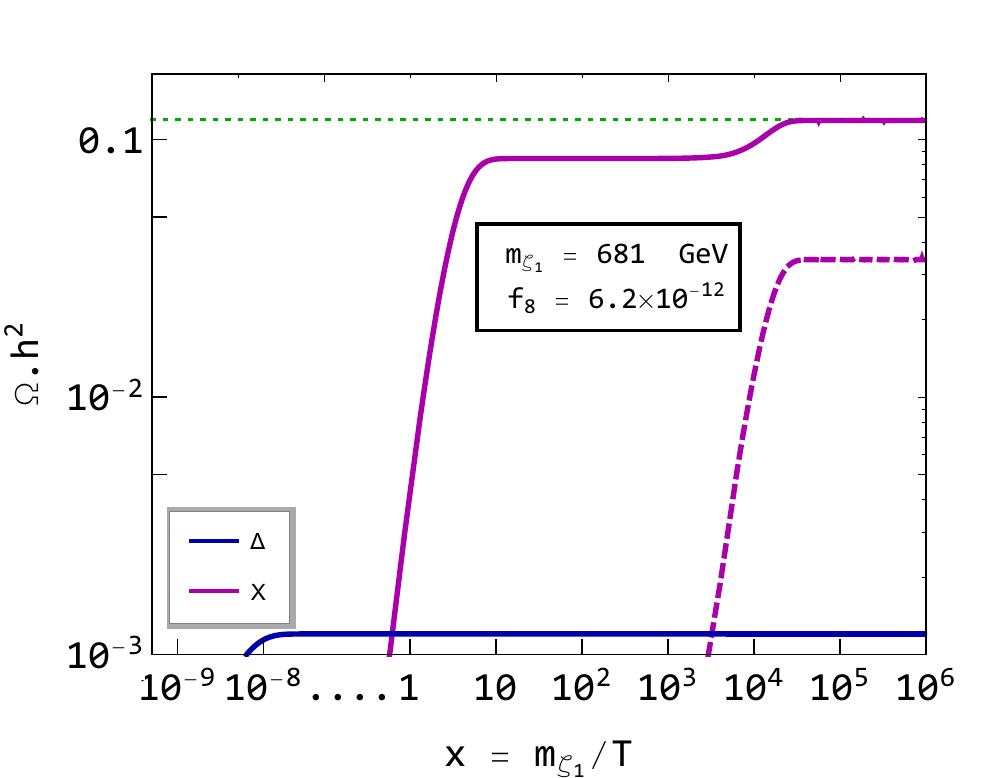}
$$
\caption{The evolution of the relic abundances $\Omega_{\Delta}\cdot h^2$ and $\Omega_{X}\cdot h^2$ as a function of  $x=M_{\zeta_1}/T$ for four different cases with $\Omega_{X}/\Omega_{\text{DM}} = \{\%25, \%50, \%75, \%99\}$ from top-left to bottom-right respectively. We have also assumed $g_N = 10^{-13}, M_{\Delta} = 10^{13}$ GeV, $M_{\zeta_2} = 200$ GeV, $M_X = 5$ GeV. The green dotted line corresponds to  $\Omega_{\Delta} + \Omega_X = \Omega_{\text{DM}}\cdot h^2 = 0.1198$, and the dashed magenta curve is the contribution just from the late $\zeta_1$ decays, i.e. after $\zeta_1$ decoupling for $x\ge x_D$. }
\label{fig:relicVSx}
\end{figure}

In Fig.~\ref{fig:relicVSx} we have plotted the evolution of the relic abundances of $\Delta$ and $X$ as a function of the dimensionless parameter $x=M_{\zeta_1}/T$. 
The parameters in Fig.~\ref{fig:relicVSx} are chosen such that $\Omega_{\Delta} + \Omega_X = \Omega_{\text{DM}}\cdot h^2 = 0.1198$.
We have chosen four different values of $m_\zeta$, to generate four different relative contributions of $X$ and $\Delta$ to the total relic density as 
$\Omega_{X}/\Omega_{\text{DM}} = \{\%25, \%50, \%75, \%99\}$ from top-left to bottom-right respectively.
The evolution of yield for $\Delta$ (shown by solid blue line in each of the plots) follows the familiar pattern of freeze-in production, which initially increases with $x$, and finally becomes 
constant as the temperature of the universe falls below the mass of $\Delta$. For $X$ (the solid magenta curve), note that the first point where the relic abundance function flattens 
is around $x\gsim 10$, where the exponential suppression inside the Bessel function becomes dominant. However, the function then rises for a second time, and captures the contribution from the decay after 
decoupling of $\zeta$. First of all, the second rise in freeze-in curve for $X$ shows that the yield from decays after the decoupling can be as large as the yield before decoupling. 
This indicates the importance of the late decay to be considered for correct evaluation of freeze-in relic density. 
The effect of late decay can be understood in more details from Eq.~\ref{eq-C1-Bol-X-7}. The share from the second integral ($x>x_D$) in Eq.~\ref{eq-C1-Bol-X-7} is at first suppressed by 
$K_1[x_D]$ and therefore it takes a long time (not until $x\sim 10^4$) for the yield to build up to an amount comparable to the yield before $x_D$. 
This happens because the asymptotic exponential suppression in $K_1(\alpha x^2 / x_D)$ 
cancels out the increasing exponential term $\exp (\alpha x^2 / x_D)$. This gain stops later as the exponential suppression from the decays becomes dominant around 
 $x \sim 10 \cdot m_{\zeta_1} / \sqrt{M_{\text{Pl}} \Gamma} \sim (5\cdot 10^4 - 10^5)
 $\footnote{Assuming $m_{\zeta_1}\sim 300$ - $700$ GeV, $m_{\zeta_2} = 200$ GeV, $m_X = 5$ GeV, we have $\Gamma \sim 10^{-23}-10^{-21}$ GeV, where $g_{\star}^{\rho} \sim 10.75$. }.
The sole contribution of the late decays to the yield are separately plotted by dashed magenta curves in each plot in Fig.~\ref{fig:relicVSx}, to show the exact $x$ for the late rise of the yield. 
It is also clear that if the decay rate was much faster, the second integral would become suppressed and we wouldn't see any effect of late production.

The behaviour of the relic abundance of $X$ (for two sample sets of values of $m_X$) as a function of $g_N$ is plotted in Fig.~\ref{fig:C1:relicVSgN}. 
We see that for small values of $g_N$, the abundance remains the same. This is because for these values of $g_N$ the decay rate is so small that the decays 
before the freeze-out of $\zeta_1$ can be neglected and the freeze-out yield of $\zeta_1$ is the same for different $g_N$ values. 
After $\zeta_1$ freezes out, all of them will eventually decay to $X$ and changing $g_N$ will only vary the time scale of these decays. 
However, beyond a threshold value of $g_N$ (characteristic to specific $m_X$), relic density rises sharply with $g_N$ as a function of $\sim g_N^2$ as the yield 
in is proportional to the decay width in case of freeze-in. This is in contrast to the freeze-out scenario where relic density is inversely proportional to annihilation cross-section. 
As we explain in the next section, smaller values of $g_N$ are ruled out by the constraints from the BBN (indicated by dashed curves in Fig.~\ref{fig:C1:relicVSgN}).

\subsubsection{Bounds on decaying relic particles from BBN and CMB} 
\label{sec:bbn}


\begin{figure}[htb!]
 $$
 \includegraphics[scale=0.52]{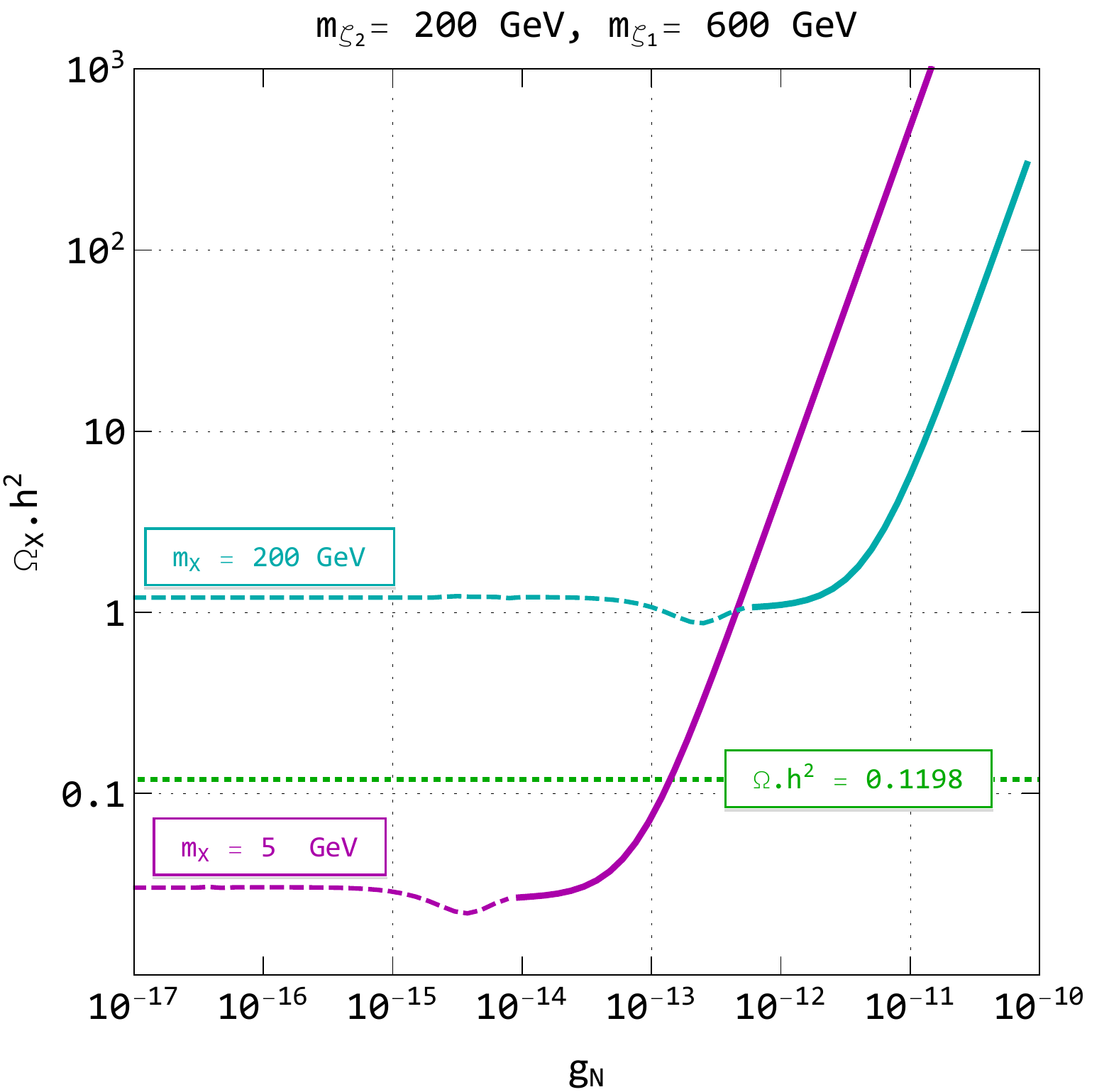}
 $$
 \caption{\label{fig:C1:relicVSgN}The relic abundance of $X$ as a function of $g_N$ for $M_{\zeta_1} = 600$ GeV, $M_{\zeta_2} = 200$ GeV. The dashed regions are ruled out from the BBN bounds. The green dotted line corresponds to $\Omega_{\text{DM}}\cdot h^2 = 0.1198$.}
 \end{figure}

Standard BBN may be significantly perturbed by the energy injections due to both neutral decays~\cite{Jedamzik:2006xz} $\zeta_1^{0} \to \overline{X} + \zeta_2^{0}$ and charged 
decays~\cite{Jedamzik:2009uy} $\zeta_1^{-} \to \overline{X} + \zeta_2^{-}$. In the case of decaying electrically charged particles with  $\tau > 100$ seconds, the existence of bound states between nuclei and the relics may significantly change nuclear reaction rates. Here, we analyze the decays with $0.05$ s $< \tau < 100$ s using the results from Ref.~\cite{Jedamzik:2006xz}, and rule out the ones with $\tau > 100$ s for simplicity. As a result, this puts a lower bound on $g_N$, depending on different choices of DM masses. The hadronic branching ratio ($B_h$) for $\zeta_1^0$ decays is $B_h^1 \approx 0.695$, and for $\zeta_1^-$ depends on parameters like $M_{n_2}, f_{\zeta}$. We can assume for heavy $n_2$  and/or small $f_{\zeta}$, the weak 
decays $\zeta_1^- \to \bar{X} + \zeta_2^- \to \bar{X} + \zeta_2^0 + W^-$ become dominant and so the hadronic branching ratio of the charged component is $B_h^2 \approx 0.9$. 
Therefore, for a conservative bound we assume hadronic branching for the both cases to be $B_h = 1$ throughout our analysis. We also assume that the mass difference $(m_{\zeta_1}^- - m_{\zeta_1}^0)$  generated during the EWSB is negligible compared to their masses before EWSB. We then extrapolate the results in~\cite{Jedamzik:2006xz} for arbitrary $m_{\zeta_1}$. The total energy of the resulting quarks in 
these decays is taken to be the same as $m_{\zeta_1}$ following~\cite{Jedamzik:2006xz}. 
We use the relation in Eq.~\ref{eq:bbnbound} to convert the bounds on the relic abundance ($\Omega^{\text{BBN}}_{\zeta_1}$) of $\zeta_1$ 
(assuming it was stable) to bounds on the DM ($X$) relic abundance (from $\zeta_1$), i.e. $\Omega_{\text{DM}}^{X}$. 
Given that $n_{\zeta_1} = n_{X}$, and $\Omega_i \propto m_i \cdot n_i$ we have:

\begin{equation}
  \Omega_{\text{DM}}^{X}\, \leq \, \left(\frac{m_X}{m_{\zeta_1}}\right) \cdot \Omega^{\text{BBN}}_{\zeta_1},
\label{eq:bbnbound}
\end{equation}

where $\Omega_{\zeta_1}^{\rm BBN}$ corresponds to the abundance of the decaying species ($\zeta_1$ in this case), 
which is fixed from BBN depending on its hadronic branching ratio and lifetime (or in other words $g_N$)~\cite{Jedamzik:2006xz}. The dashed portions in 
Fig.~\ref{fig:C1:relicVSgN} are thus ruled out by the BBN bound since smaller values of $g_N$ are discarded in order to prevent prolonged decays that may disturb the standard BBN mechanism.

A decaying long-lived DM candidate can be constrained by various observations. For example, it can alter the ionization and heating history of the CMB and its power spectrum~\cite{Mambrini:2015sia}. Decays to several decay modes {\it e.g.} $b\bar{b}$, $WW$, $\mu\bar{\mu}$ can also be constrained by the AMS-02 precise measurements of the antiproton/proton ($p/\bar{p}$ ) fraction~\cite{Aguilar:2016kjl}, as no evidence of new source of antiproton has been found in these data. This in turn, results in the following bounds on the hadronic decay life time of the DM~\cite{Mambrini:2015sia}:

\begin{align}\label{eq:bound}
\text{CMB:} \qquad \tau(\text{DM} \to b\bar{b}) \gtrapprox  \, 10^{24} \, s, \\
\text{AMS-02:} \qquad \tau(\text{DM} \to b\bar{b}) \gtrapprox \, 10^{27} \, s.
\end{align}


As we already have shown, the $\Delta$'s have lifetime comparable to that of the universe. Now, if we set $m_X=5~\rm GeV$, $g_N=10^{-12}$ and $f_8=10^{-11}$ then for $u_3=100~\rm GeV$ 
we obtain the following decay width and corresponding decay lifetime for the $\Delta$'s:

\begin{align}
\Gamma\left(\Delta_1\to \bar{X}\bar{X} b\bar{b}\right)=& 4.1\times 10^{-60}~\rm GeV  \quad \longrightarrow \quad \tau_{\Delta_1}\sim 10^{36}~\rm sec\\
\Gamma\left(\Delta_2\to \bar{X} b\bar{b}\right)=&  5.1\times 10^{-59}~\rm GeV \quad \longrightarrow \quad \tau_{\Delta_2}\sim 10^{35}~\rm sec\\
\Gamma\left(\Delta_3\to b\bar{b}\right)=&  1.2\times 10^{-58}~\rm GeV \quad \longrightarrow \quad \tau_{\Delta_3}\sim 10^{34}~\rm sec
\end{align}

All these decay rates turn out to be much longer than the bounds mentioned in Eq.~\ref{eq:bound}, and so only the bounds from the $\zeta_1$ decays are relevant for final estimate of allowed parameter space. 
Note that, since some of the energy of $\Delta_1$ goes into $\bar{X}$'s, the actual bounds are smaller, and these numbers are conservative.

\subsubsection{Summary of available parameter space} 
\label{sec:bbn}

\begin{figure}[htb!]
 $$
 \includegraphics[scale=0.62]{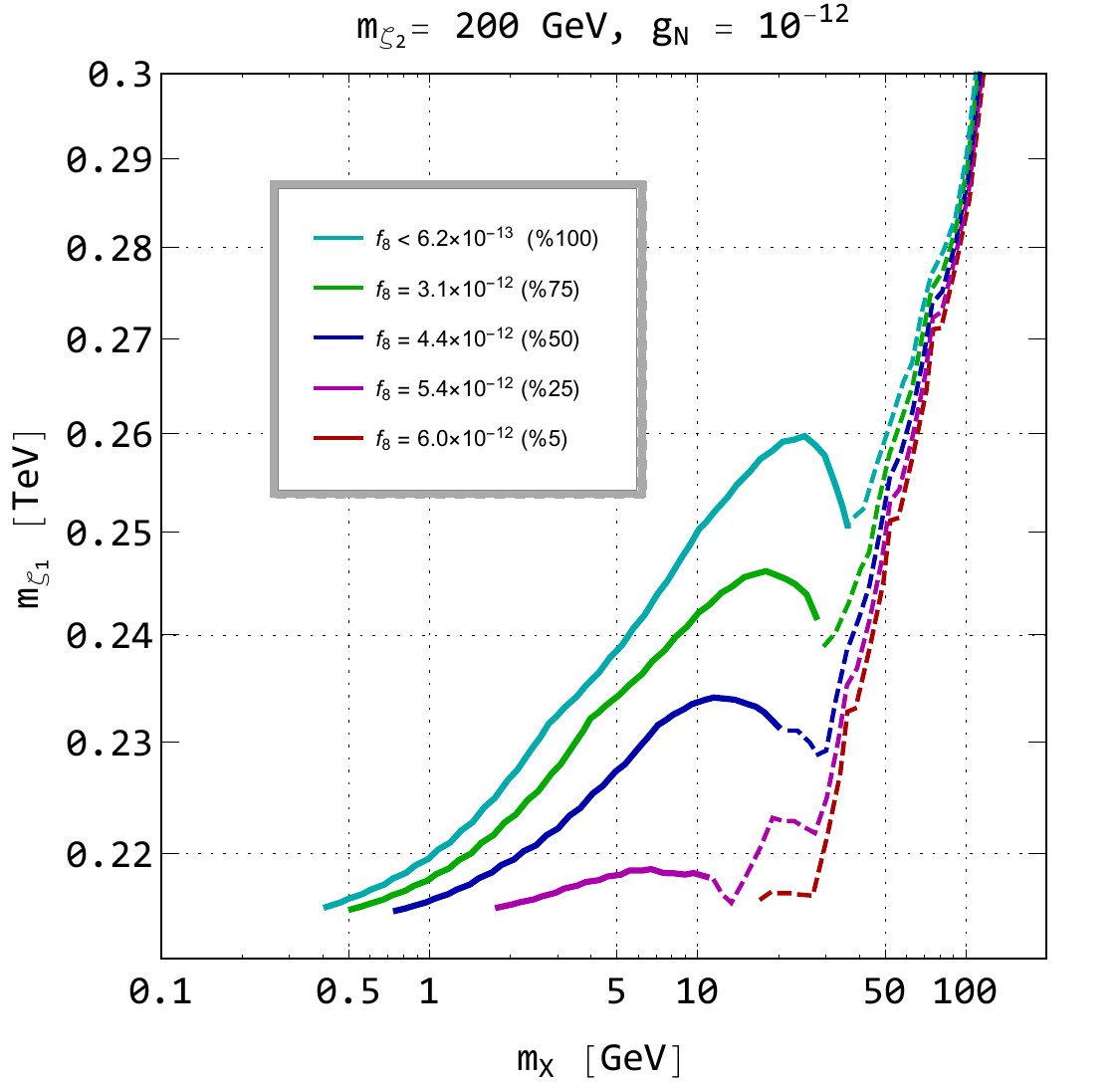}
 \includegraphics[scale=0.6]{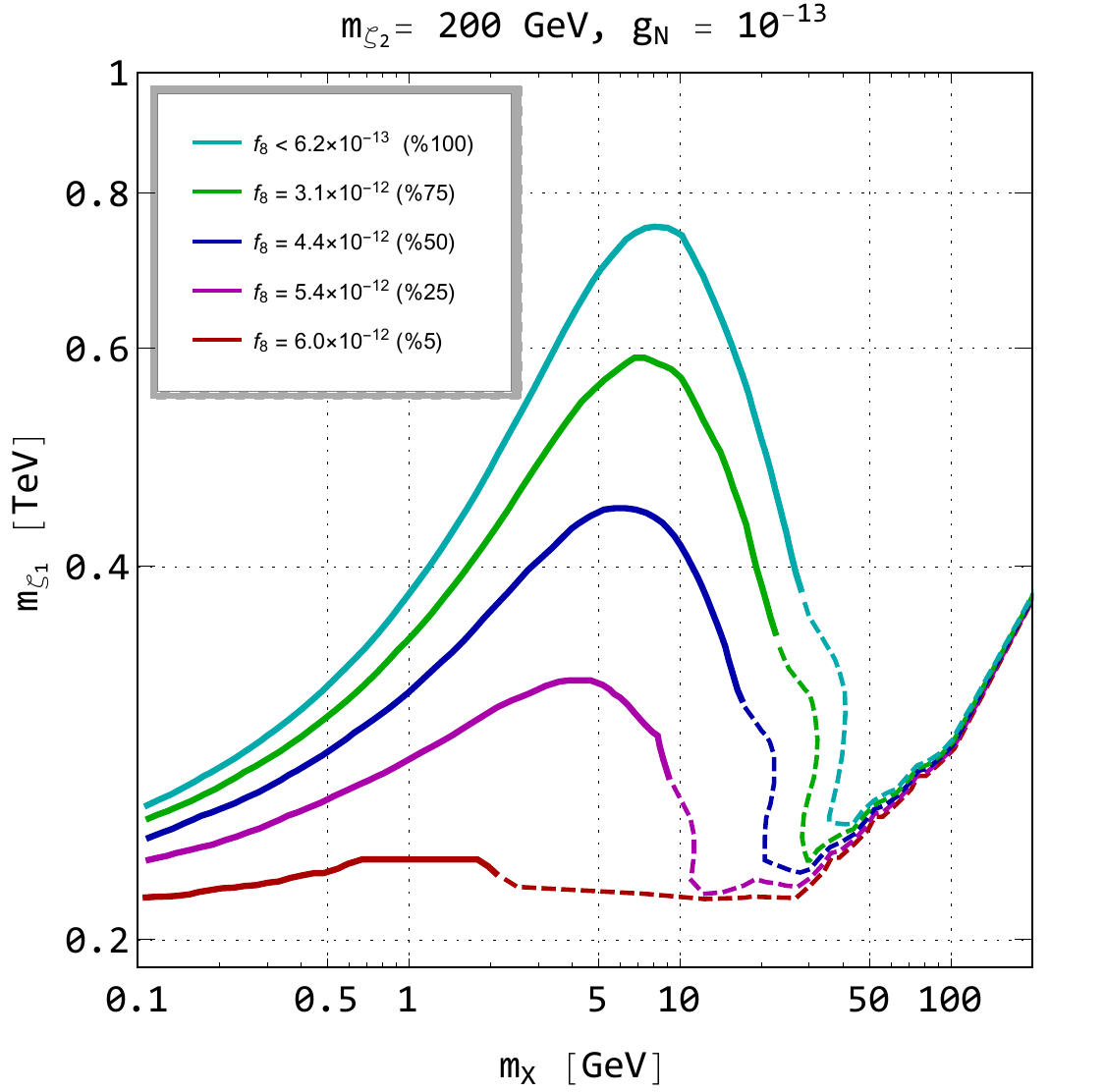}
 $$
 $$
\includegraphics[scale=0.6]{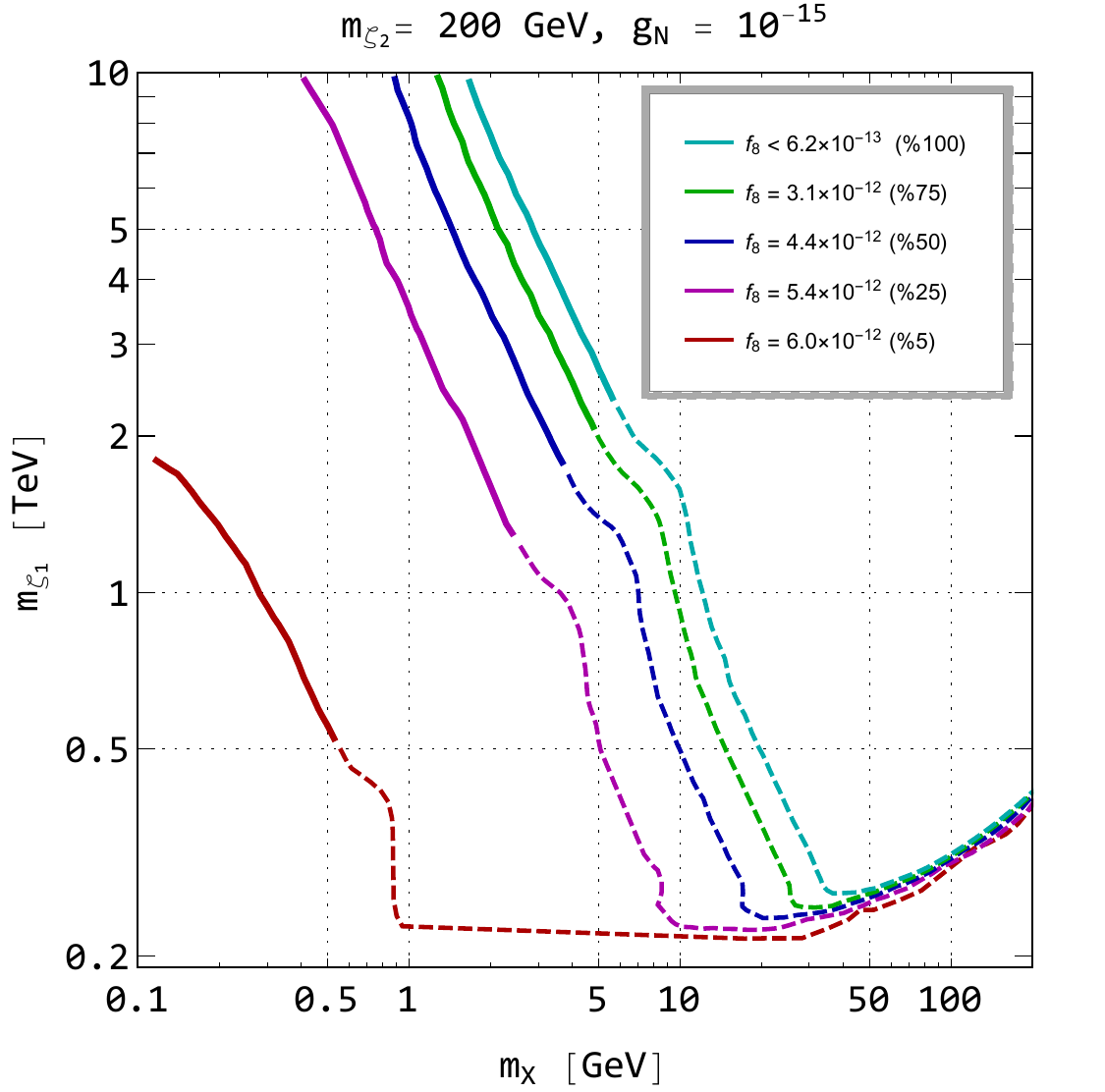}
\includegraphics[scale=0.6]{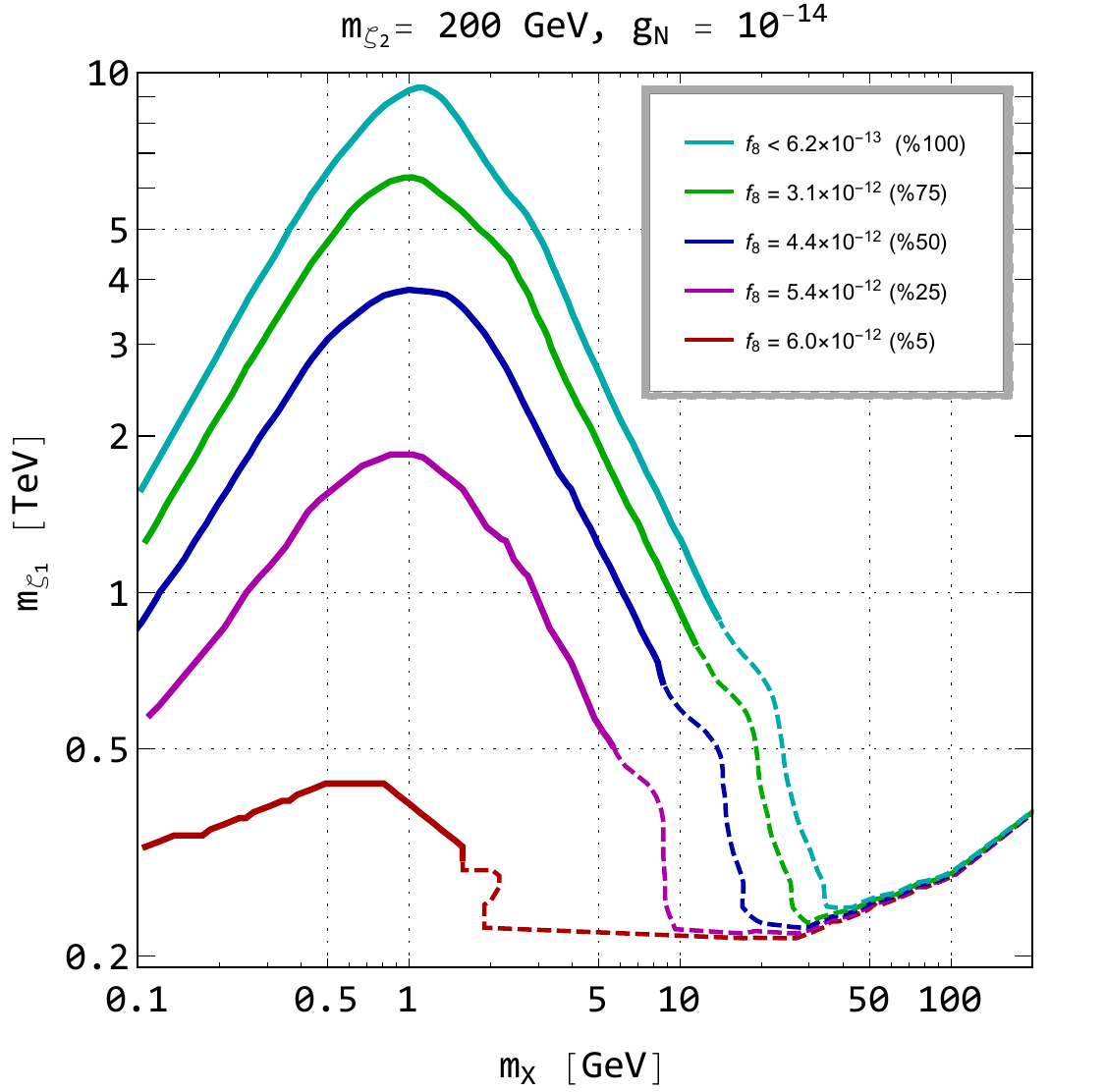}
$$
\caption{\label{fig:C1:paramspace} Parameter space of the model satisfying relic abundance constraint in $m_{\zeta_1}$-$m_X$ plane when $g_N=\{10^{-12},10^{-13},10^{-14},10^{-15}\}$ (clockwise from top left) and $M_{\zeta_2}=200~\rm GeV$ are kept fixed for each plot. Each contour satisfies total relic abundance of $\Omega_{\text{DM}}\cdot h^2 = 0.1198$ for different choices of the scalar quartic coupling $f_8$. The numbers inside the parentheses are the percentage of DM comprised of $X$. The dashed portion in each curve is ruled out by the BBN constraints.}
 \end{figure}

\begin{figure}[htb!]
\centering
\includegraphics[scale=0.72]{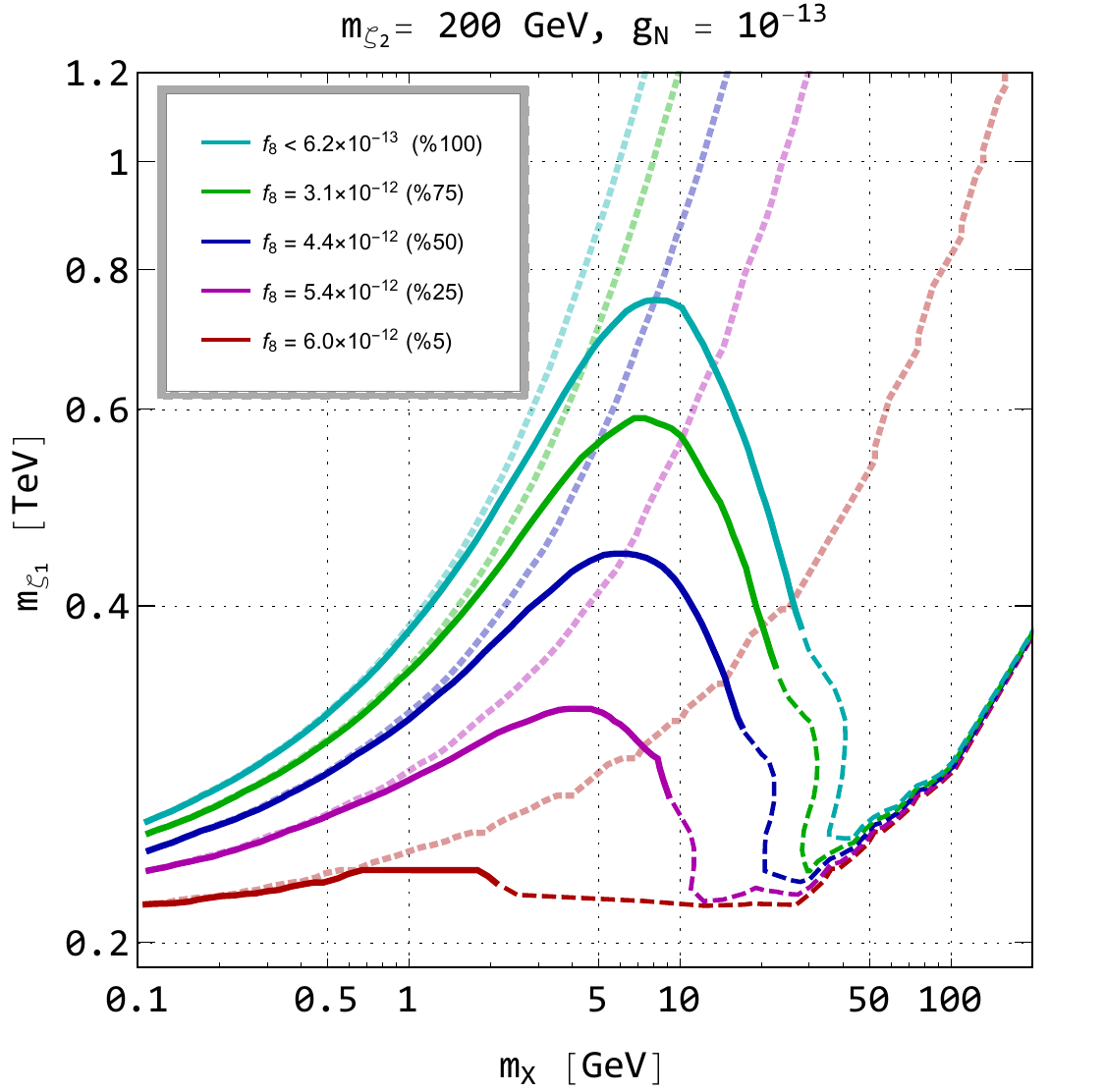}
\caption{\label{fig:C1:paramspacecomp}Parameter space of satisfying the relic abundance constraint for $g_N = 10^{-13}$ in $m_{\zeta_1}$-$m_X$ plane. The opaque dotted curves in the background are the corresponding contours if the late decays are ignored. The dashed regions are ruled out by the BBN constraints. }
\end{figure}

The relic density allowed parameter space for $X$ is plotted in $m_{\zeta_1}-m_X$ plane in Fig.~\ref{fig:C1:paramspace} for different choices of $g_N=\{10^{-12},10^{-13},10^{-14},10^{-15}\}$ (clockwise from top left). 
Each plot contains several contours corresponding to a specific percentage of the total DM density coming from $X$ (numbers inside parenthesis). Therefore, each curve corresponds  to a specific 
value of $f_8$ such that the rest of the DM abundance ($100\%-$ the percentage in the parenthesis) is coming from $\Delta$. This is possible because the 
relic density of $X$ and that of $\Delta$ are uncorrelated in the model. We see from the plots that constant $\Omega_X$ curves rises in $m_{\zeta_1}-m_X$ plane for larger $m_X$ and then takes a 
sharp turn to decrease the required $m_{\zeta_1}$ to satisfy relic density and then rises again. In order to interpret the shape of the contours in Fig.~\ref{fig:C1:paramspace}, 
let's first take a look at the parameter space assuming the late decays are ignored. This is shown in Fig.~\ref{fig:C1:paramspacecomp} with the dashed upward moving faint lines in the background. These 
contours basically correspond to constant relic density ($\Omega_Xh^2$), which is mainly dictated by $\Omega_X \propto c \approx m_X \cdot \Sigma_{\zeta}$, since the second integral in Eq.~\eqref{eq-C1-Bol-X-8} is to be ignored. 
Therefore, it is straightforward to see that the larger is the value of $m_X$, the smaller $\Sigma_{\zeta}$ has to be and therefore $m_\zeta$ will needs to be larger as $\Sigma_{\zeta} \propto 1/m_\zeta^2$.
It is also clear, for small $m_X$ (as we require larger $\Sigma_{\zeta}$) the second integral in Eq.~\eqref{eq-C1-Bol-X-8} has smaller contribution due to exponential suppression from $\Sigma_{\zeta}$. This is reflected in Fig.~\ref{fig:C1:paramspacecomp} where the case with late decays and the case without them merge together at low $m_X$.
As we trace these contours from low $m_X$ region (left) to high $m_X$ regions (right), we observe a departure for the curves where the late decays are involved.
Now, with larger $m_X$, $\Sigma_{\zeta}$ decreases. This increases the {contribution from the late decays as the second integral in Eq.~\eqref{eq-C1-Bol-X-8} is proportional to $e^{-\Sigma_{\zeta}}$}. Therefore, if we compare the case with late decay ($\Sigma_{\zeta}^1$) to that without late decay ($\Sigma_{\zeta}^2$), it is easy to see that:
\bea
\Sigma_{\zeta}^1>\Sigma_{\zeta}^2;\nonumber \\
\implies m_{\zeta_1}^1<m_{\zeta_1}^2,
\eea
which is reflected in the drop of the curve beyond certain $m_X$.

We can estimate the value of $m_X$ for which this change of behaviour happens. As mentioned at the end of subsection~\ref{sec:beq}, the exponential suppression in the second 
integral becomes notable around $x_e \sim m_{\zeta_1} / \sqrt{M_{\rm Pl}\Gamma} = 1/\sqrt{\Sigma_{\zeta}}$. Since the yield from the second integral becomes relevant around $x_{obs} \sim 10^{4}$ (see Fig.\ref{fig:C1:paramspace}), 
we can use $x_e \lsim x_{obs}$ as a criterion for this change. Let's start from low $m_X$ region and trace the constant $c$ contours for which $\Omega_{X}\cdot h^2 \approx 0.1198$. 
Using Eq.~\ref{eq-C1-Bol-X-8} and~\ref{eq-C1-Omega-Tot} we obtain $c \approx 10^{-7}$ for the cyan dashed curve where 100 \% contribution to the relic abundance comes from $X$. 
Now, since $x_e \approx \sqrt{M_X/c}$, this criterion transmits to $ M_X \lsim 10$ GeV. If we choose the curve for $5\%$ yield instead, we get $c\sim 10^{-8}$ and $ M_X \lsim 1$ GeV. 
This explains the sharp turn in all of the curves that happens around $M_X \approx 1-10$ GeV. Also note that as $m_{\zeta_1}$ is lowered and $m_X$ is increased, 
we get into a region where the decay rate is so slow that it violates the BBN bounds (dashed region). As $m_X$ is increased even further, all the contours converge to the 
kinematically forbidden boundary i.e. $M_{\zeta_1} = M_X + M_{\zeta_2}$. This characteristic feature is reflected in each plot in Fig.~\ref{fig:C1:paramspace} 
for different choices of $g_N$ with $M_{\zeta_2}$ fixed at 200 GeV.


%

\subsection{Degenerate $\Delta$'s with $2 m_{X} < m_{\Delta}$ and $f_8 \sim \mathcal{O}(1)$}

In this case let us first compute the lifetime of the $\Delta$'s assuming $f_8\sim\mathcal{O}(1)$.
We once again recall here that $m_{\Delta}\sim\mathcal{O}(10^{13})~\rm GeV$. 
Let us choose $m_X=5~\rm GeV$ and $g_N\sim\mathcal{O}(10^{-13})$ for which $X$ yields relic density in correct ballpark.
We then find, {$\Gamma_{\Delta_3}\to t\bar{t}=1.3\times 10^{-35}~\rm GeV$ which dubs into a lifetime of $\sim 5\times 10^{11}~\rm sec \ll \tau_{\text{universe}}$.} 
With the same set of parameters, we find {$\Gamma_{\Delta_2\to\overline{X}t\overline{t}}=5.5\times 10^{-36}~\rm GeV$,
which gives rise to a lifetime of $\sim 10^{11}~\rm sec$.} 
The decay width for $\Delta_1$ turns out ot be {$\Gamma_{\Delta_1\to\overline{X}\overline{X}t\bar t}=4.5\times 10^{-37}~\rm GeV$, with a lifetime of $\sim 10^{12}~\rm sec$.} Therefore, with $f_8\sim\mathcal{O}(1)$ we end up with a situation
where all of the triplet scalar components are unstable and
don't contribute to the DM relic abundance. Therefore we end up with a single component $X$ DM set up. The initial freeze-out abundance of $\Delta$'s (before they decay to $X$) is determined by 
\begin{equation}
\sigma(\Delta_i \Delta_i^*\to \Phi \Phi^* )\cdot v_{\rm rel} = \frac{f_8^2}{32 \pi m_{\Delta_i}^2} \approx 10^{-28}\cdot f_8^2,
\end{equation}
which yields a freeze-out abundance for $\Delta$ at the order of: 
\begin{equation}
\Omega_{\Delta}^{\rm FO}\cdot h^2 \approx 10^{18}
\end{equation}
If we use the fact that the number of $X$'s produced from the decays of $\Delta$, i.e. $n_{X}^{\rm FI} \sim n_{\Delta}^{\rm FO}$, we then see that $\Omega_{X}^{\rm FI} = (m_{X}/m_{\Delta})\Omega_{\Delta}^{\rm FO} $ is about $6$ orders of magnitude higher than the correct DM abundance, which rules out this scenario with $f_8 \sim \mathcal{O}(1)$. 

\section{Collider search}
\label{sec:collider}
Possible collider signatures of this model in the context of usual freeze-out of $X$ was elaborated in~\cite{Barman:2018esi}. It was pointed out that, the only field connecting the dark sector (having $SU(2)_N$ charge) 
with the SM is the scalar bi-doublet. Therefore, the charged and neutral components of the scalar bi-doublet can be produced at the LHC via the diagrams shown in the top panel of Fig.~\ref{fig:collider}. We should also note that the other scalars, namely the $SU(2)_N$ doublet ($\chi$) and triplet ($\Delta$) possess Higgs portal interactions. However, we have assumed all those portal couplings to be small to satisfy the Higgs data (see Sec.~\ref{sec:model}). In particular, the Higgs portal coupling ($f_8$) between the triplet ($\Delta$) 
and SM doublet ($\phi$) was assumed to be small to address correct freeze-in production of the triplet components, which serve as DM. Therefore, the other particles do not have any chance of being produced at the LHC.
The variation of production cross-section of the charged and neutral components of the scalar bi-doublet at the LHC with respect to its mass $m_{\zeta_1}$ for $E_{\text{CM}}=$14 TeV is shown in the bottom panel of Fig.~\ref{fig:collider}. It should be noted that charge current contribution is more pronounced than the neutral current contribution which we pointed out in our earlier analysis~\cite{Barman:2018esi}. We have implemented the model in {\tt CalcHEP}~\cite{Belyaev:2012qa} and
used {\tt CTEQ6l}~\cite{Placakyte:2011az} as a representative parton distribution function for generating this process. The SM gauge coupling is mostly responsible for the production of the bi-doublet components at the LHC. The Higgs portal 
interactions are again assumed to be small to be compatible with the Higgs data.

\begin{figure}[htb!]
$$
\includegraphics[scale=0.42]{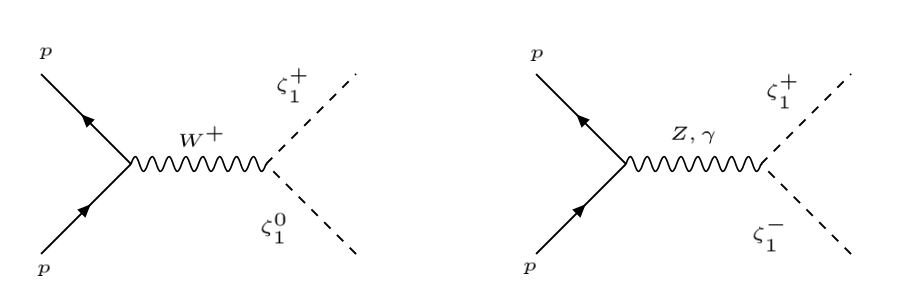}
$$
$$
\includegraphics[scale=0.4]{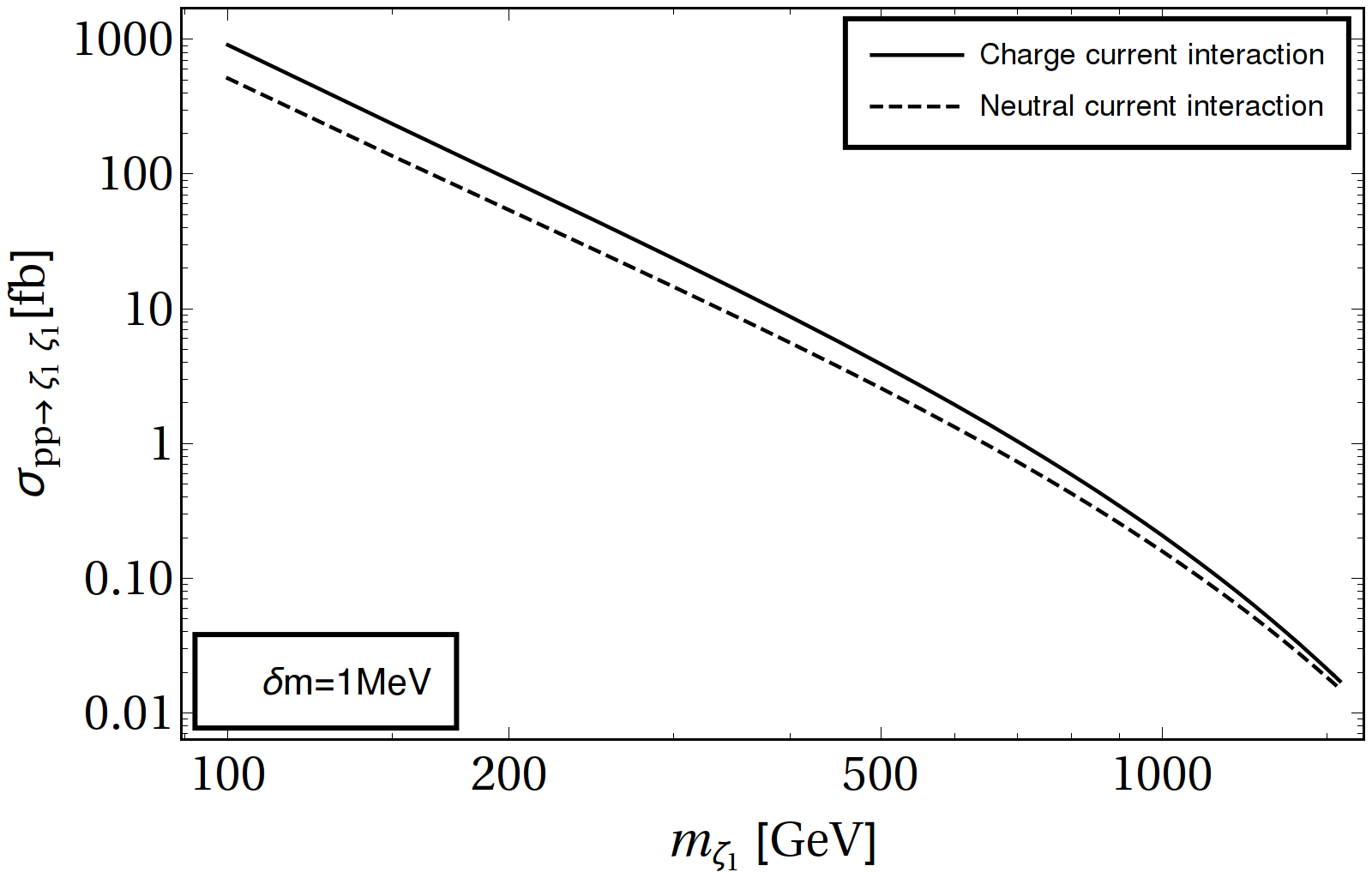}
$$
\caption{Top: Associated and pair production of heavy charged scalars at the LHC via charged current (LHS) and neutral current (RHS) interaction. Bottom: Variation  of production cross section of $pp\to \zeta_1^{\pm}\zeta_1^{0},\zeta_1^{+}\zeta_1^{-}$ at the LHC with $E_{\text{CM}}=$14 TeV. Contributions are from charged current (\textbf{solid}) and neutral current (\textbf{dashed}) interactions for $\delta m=1~\rm MeV$.}
\label{fig:collider}
\end{figure}

If the additional neutrinos ($n_1,n_2$) are lighter than
$\zeta_1^{\pm,0}$ (which is quite legitimate in WIMP scenario even after addressing correct neutrino mass generation), 
then these bi-doublet scalars can further decay to neutrinos and SM leptons ( $\zeta_1^+ \to \ell^+ + n_{1R}$) via Yukawa interaction given in Eq.~\ref{eq:yukawaint}, 
giving rise to the following signatures in colliders~\cite{Barman:2018esi}:  

\begin{itemize}
 \item Single lepton with missing energy ($1\ell^{\pm}+\slashed{E}_T$) due to charged current interaction.
 \item Opposite sign di-lepton with missing energy $(\ell^+\ell^-+\slashed{E}_T)$ due to neutral current interaction.
\end{itemize}

In the WIMP-like freeze-out scenario of $X$, therefore this model may leave an imprint of leptonic signal excess at the colliders. However, in the present framework, we assume that the DM production is from the decay of $\zeta_1 \to \zeta_2+X$, and the additional neutrinos ($n_1,n_2$) to be heavier than $\zeta_1$. It is understandable that the DM analysis will be modified by the corresponding decay branching ratio if $n_{1,2}$ are lighter than $\zeta_1$. 
It is also instructive to remind that heavier $n_{1,2}$ ($\sim 10^6$ GeV) arises naturally when we assume the Yukawa in a legitimate ball park $f_\Delta \sim 0.1$ via a large 
$u_3 \sim 100$ GeV to address the required neutrino mass (see Sec.~\ref{sec:numass}). However, one may still assume the presence of lighter  $n_{1,2}$
($\sim$ TeV) with fine tuned $f_\Delta \sim 10^{-7}$ and $m_{n_{1R}}<m_{\zeta_1}$, yielding single lepton and di-lepton signature as before. 
However in that case, the DM production will be further suppressed and 
appropriate decay branching ratios have to be assumed. 

Therefore, in the light of DM analysis performed here, the neutrinos ($n_{1,2}$) are heavier than the scalar bi-doublet
$\zeta_{1,2}^{\pm},\zeta_{1,2}^0$. As a result, these scalars can not decay
to RHN plus SM leptons. Due to radiative correction, there should be a small mass splitting between the charged and neutral
components of the bi-doublet: $\delta m$. As a result, the charged 
scalars, once produced, can undergo the decays shown in 
Fig.~\ref{fig:decay} via off-shell $W$ or via off-shell $n_{1,2R}$ to $\zeta_1^0+\ell+\nu_\ell$. 

\begin{figure}[htb!]
$$
\includegraphics[scale=0.5]{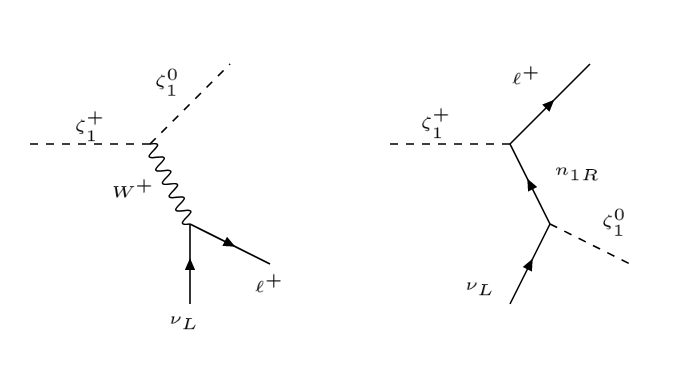}
$$
\caption{Decay of the charged components of the scalar bi-doublet
 which is kinematically possible since $m_{\zeta_1^+}-m_{\zeta_1^0} \equiv \delta m> 0$.}
\label{fig:decay}
\end{figure}

\begin{table}[htb!]
\begin{center}
\scalebox{1.0}{
\begin{tabular}{|c|c|c|c|c|c|c|c|c|}
\hline
$\delta m$ (GeV) & $Br(\zeta_1^{\pm}\to\zeta_1^0,\ell^{\pm},\nu_L)$ (via $W$) & $Br(\zeta_1^{\pm}\to\zeta_1^0,\ell^{\pm},\nu_L)$ (via $n_{1R}$) \\
\hline\hline 
0.1 & 0.99 & 0.0007    \\
0.5 & 0.99 & 0.0007 \\
\hline
\end{tabular}
}
\end{center}
\caption {Decay branching ratio of $\zeta_1^{\pm}\to\zeta_1^0,\ell^{\pm}$ via $W$ and via $n_{1R}$. Here we have chosen $m_{\zeta_1}=1.2~\rm TeV$ and $\delta m=100~\rm MeV$ with RHNs of mass $M=10^5~\rm GeV$.}
\label{tab:br-zetapl}
\end{table} 

\begin{table}[htb!]
\begin{center}
\scalebox{1.0}{
\begin{tabular}{|c|c|c|c|c|c|c|c|c|}
\hline
$\delta m$ (GeV) & $\Gamma$ (GeV) & $L=c\tau$ (m) \\
\hline\hline 
0.1 & $3.12\times 10^{-18}$ & 63.05    \\
0.3 & $7.58\times 10^{-16}$ & 0.26   \\
0.5 & $9.76\times 10^{-15}$ & 0.20 \\
\hline
\end{tabular}
}
\end{center}
\caption {Decay lengths for the three-body decay of $\zeta_1^{\pm}$ assuming maximum the radiative mass splitting $\delta m$ of 500 MeV for $m_{\zeta_1^{\pm}}=1.2~\rm TeV$, RHN mass $M=10^5~\rm GeV$ and $f_{\zeta}\sim\mathcal{O}(1)$.}
\label{tab:track}
\end{table}

\begin{figure}[htb!]
$$
\includegraphics[scale=0.38]{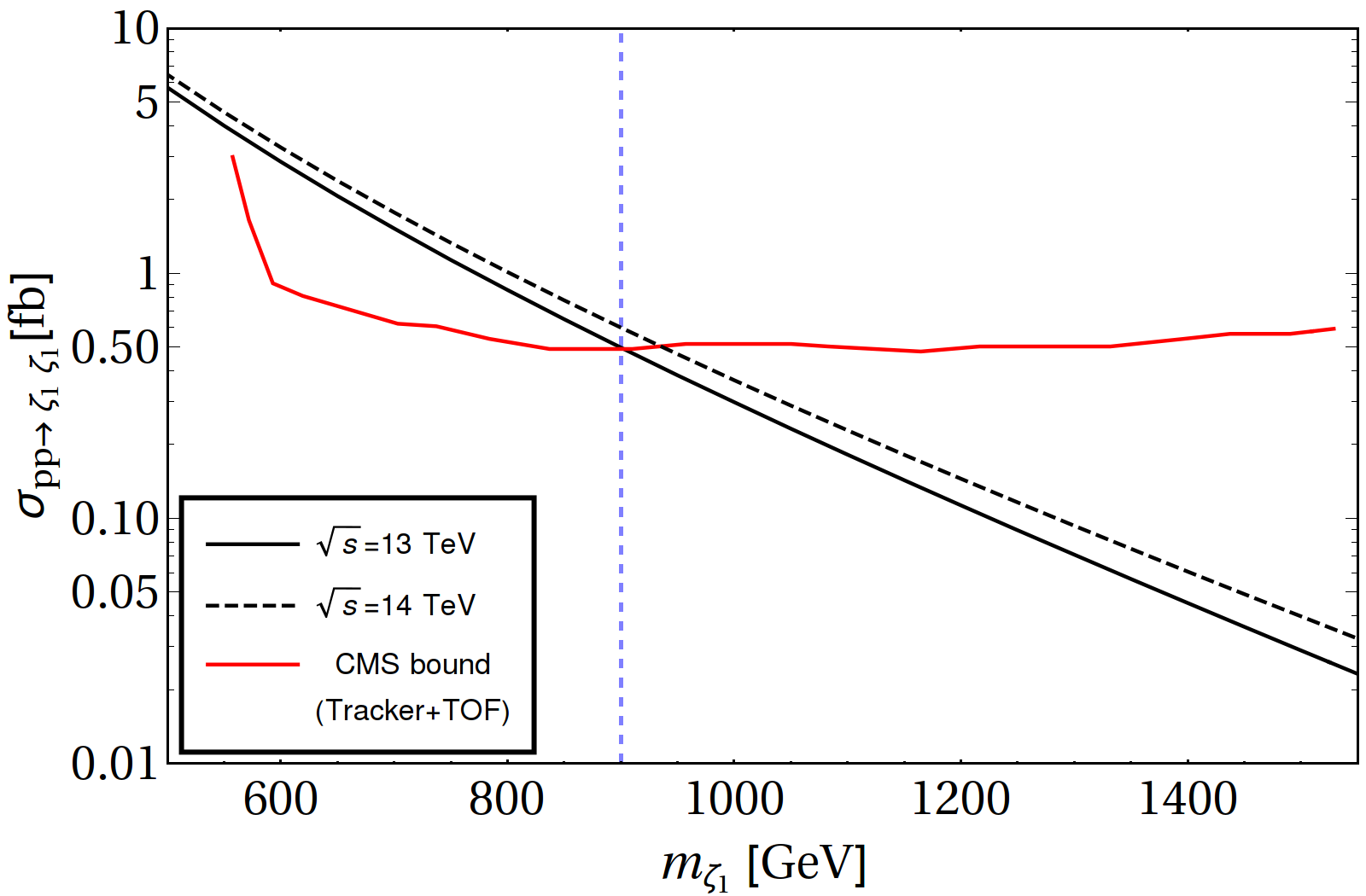}
\includegraphics[scale=0.38]{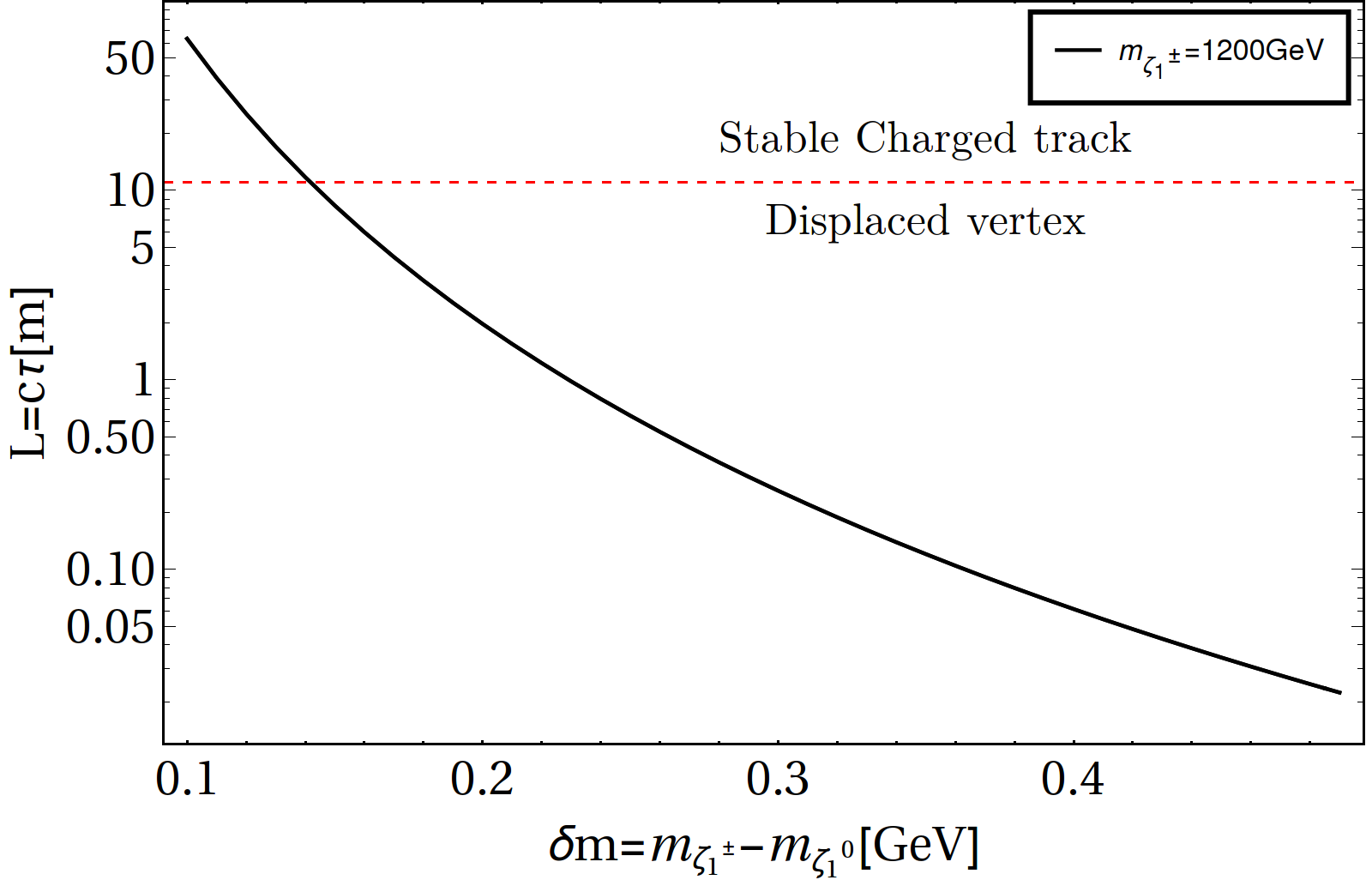}
$$
\caption{Left: The variation of production cross section of $pp\to \zeta_1^{\pm}\zeta_1^{0},\zeta_1^{+}\zeta_1^{-}$ at the LHC with $E_{\text{CM}}=$13 TeV (solid black curve) and $E_{\text{CM}}=$14 TeV (dashed black curve). The red solid curve shows the bound from CMS searches for HSCP at $\sqrt{s}=13~\rm TeV$ and luminosity $12.9~\rm fb^{-1}$. The vertical dashed line shows the minimum mass that is allowed by the CMS exclusion limit; Right: Decay length of 
$\zeta_1^{+} \to \zeta_1^{0}+\ell^-+\bar{\nu}_\ell$ (in meter) in terms of mass splitting $\delta m \sim\mathcal{O}(100~\rm MeV)$ (in GeV). We show the possibilities of displaced vertex 
or stable charged track segregated via dashed red line at collider assuming a legitimate $m_{\zeta_1^{\pm}}=1.2~\rm TeV$ .}
\label{fig:track}
\end{figure}

Note that the diagram in the LHS of Fig.~\ref{fig:decay} involves the vertex $\zeta_1^+\zeta_1^0W$, which is proportional to $g_L(p_1+p_2)_{\mu}$ where $p_1$ and $p_2$ are the momenta of the incoming and outgoing scalars, 
while the other vertex involves only the SM gauge coupling $g_L$. The diagram on the RHS of Fig.~\ref{fig:decay}, on the other hand, only depends on the Yukawa coupling $f_{\zeta}$, 
which we assume to be $\sim\mathcal{O}(1)$. Since $n_{1R}$ is heavy, these decays are dominated by the $W$-mediated process as in the LHS of Fig.~\ref{fig:decay}. 
In Tab.~\ref{tab:br-zetapl}, we show a couple of sample points with two different mass splitting, where 99\% of the branching is carried away by the $W$-mediated decay as $m_W\ll m_{n_{1R}}$. Also note that in the limit 
$\delta m \ll m_{\zeta_1}$, the decay width mostly depends on $\delta m$ (which controls the phase space) and not on $m_{\zeta_1^{\pm}}$. For example, if we fix $\delta m=0.1~\rm GeV$, then for 
$m_{\zeta_1^{\pm}}=1.2~\rm TeV$, the decay width is $3.127\times 10^{-18}$ GeV, which changes to $3.126\times 10^{-18}$ GeV for $m_{\zeta_1^{\pm}}=1.5~\rm TeV$. 
For $\delta m\sim\mathcal{O}(100~\rm MeV)$ this model can give rise to charge track and/or displaced vertex~\cite{Co:2015pka,Chakraborti:2019ohe} at the colliders that can be probed by 
current or future experiments~\cite{Lubatti:2019vkf,Mavromatos:2016ykh}. A heavy stable charged particle (HSCP) such as $\zeta_1^+$ in our model, will typically travel with a velocity $\beta\equiv\frac{v}{c}<1$. 
Hence, as it passes through the silicon detectors, it produces an ionizing track with higher ionization energy loss rate (larger $dE/dx$) compared to the SM particles~\cite{Belanger:2018sti}. 
Also, if the HSCP decays outside the detectors, the time of flight (TOF) measured by the muon system will be longer than that of relativistic muons. These two features can distinguish non-standard HSCPs from the SM particles. 

Typically, for $c\tau\lsim\mathcal{O}(10~\rm m)$ the searches for HSCPs are done via displaced vertex signatures, while for $c\tau>\mathcal{O}(10~\rm m)$ a sizable fraction will decay only after crossing the
 tracker and/or muon chamber~\cite{Belanger:2018sti}. We use the CMS bound on HSCP production cross-section at the LHC~\cite{CMS:2016ybj} to constrain the mass of $\zeta_1$ in our model. 
The $\zeta_1$  production cross-section is plotted in the LHS of Fig.~\ref{fig:track} with respect to $\zeta_1$ mass. We have also shown the limit from the CMS tracker+TOF analysis~\cite{CMS:2016ybj} 
(for luminosity of $12.9~\rm fb^{-1}$) with a solid red curve. As one can see, these constraints rule out $m_{\zeta_1^{\pm}} \lsim 1~\rm TeV$. We have tabulated the decay lengths for some selected 
values of $\delta m$ with $m_{\zeta_1^{\pm}}=1.2~\rm TeV$ in Tab.~\ref{tab:track}. The same is also shown in RHS of Fig.~\ref{fig:track} where we have plotted the variation of the decay length with respect to 
mass splitting $\delta m$ for $m_{\zeta_1^{\pm}}=1.2~\rm TeV$. As mentioned, the decay length decreases as the mass splitting $\delta m$ increases. We have indicated a red dashed line, 
above which ($c\tau>10~\rm m$) the model can give rise to stable charged track and below ($c\tau\lsim10~\rm m$) displaced vertex signature.

%
\section{Summary and Conclusions}
\label{sec:concl}

In this draft, we have analysed FIMP realization of a non-abelian vector boson DM in $SU(2)_N$ extension of the SM. Non-abelian cases are important for several reasons, one because they require {\it non-minimal} extensions in the scalar sector 
for spontaneous symmetry breaking of the additional $SU(2)_N$ and therefore serve as an important framework to elaborate on Higgs physics in light of the present data. The model at hand also addresses neutrino mass 
generation and therefore neutrino mass constraint plays an important role in identifying the allowed parameters of the model together with DM constraints. For example, this exercise has led us to conclude 
(i) scalar triplets are super heavy (of the scale of $SU(2)_N$ breaking), (ii) sterile neutrinos assumed in the model can also be naturally heavier than scalar bi-doublet. 
Now, both of these two conclusions have immense phenomenological consequence. Therefore, the exercise performed in this analysis can serve as a benchmark to address non-thermal DM production together with neutrino mass and Higgs phenomenology.  

One of the important outcomes of this analysis turns out to be out-of-equilibrium decay of a heavier particle to DM. In the present context, scalar bi-doublet $\zeta_1^{0,\pm}$ decays into its lighter partner $\zeta_2^{0,\pm}$ plus DM $X$, and this is solely responsible for non-thermal production of the DM. The production of DM occurs after $SU(2)_N$ symmetry breaking ($\sim 10^{12}$ GeV) and before electroweak symmetry breaking. We find that, the decay of $\zeta_1^{0,\pm}$ provides a significant contribution to DM relic density even after the freeze-out of $\zeta_1^{0,\pm}$ from thermal bath.  This results in a sharp deviation of relic density contour in $M_{\zeta_1}-M_X$ allowed plane compared to the case where the late productions are neglected. The impact of this conclusion can be made in a generic and model independent way, to demand that any particle in thermal bath whose decay is slow enough ($\sim 10^{-24}$ GeV) can contribute significantly after freeze out and alter the available parameter space to a significant extent. We provide with a generic expression
for the DM yield including the late decays that may serve useful in identifying such contributions for any model.  

It is also important to note the connection between the dark sector and neutrino sector addressed in this model. The requirement of having a freeze-in vector boson DM makes the $SU(2)_N$ scalar triplet ($\Delta$), assumed for 
neutrino mass generation through inverse seesaw, superheavy ($\sim 10^{12}$ GeV). The decay modes of the neutral component of the triplet (to $t \bar{t}$ or to $\nu\bar\nu$) turns out to be extremely small, 
thanks to small $g_N~\rm{and}~ f_8$ couplings. Therefore they are stable and serve as additional DM components in the model. It is intriguing to note that the correct relic density (or under abundance) for 
$\Delta$ can only be addressed if they are also produced non-thermally through Higgs quartic interaction. CMB data constrains the decay life time of long lived DM particles to hadronic final states ($\Delta \to b\bar{b} $ in our case) 
to be greater than $\tau(\text{DM} \to b\bar{b}) \gtrapprox  \, 10^{24} \, s$. However, since the correct relic density is achieved for $\tau_{\Delta} \gtrapprox 10^{36} \, s$, this bound does not affect that part of the parameter space where $\Delta$ is a viable DM candidate.


There are other constraints as well. For example, bounds from AMS-02 constrains life time of hadronically decaying DM (again $\Delta \to b\bar{b} $ in our case) 
while BBN data constrains life time of semi-stable hadronically decaying charged and neutral particle ($\zeta_{1,2}^{0,\pm}$ in our case). 
The last bound crucially tames down a large allowed parameter space of our model by ruling out DM masses above $\sim 50$ GeV and $g_N \lsim 10^{-14}$. 
Constraints from CMB and AMS-02 turn out to be less sensitive due to the very long life time of scalar triplet DM. 

The same model has been studied for WIMP realization as mentioned before. It is therefore important to identify the difference in their phenomenological implication. While freeze-in makes the DM 
insensitive to direct search, the WIMP can be detected via future direct search experiments. It is important to note that WIMP-like $X$ is allowed upto $\sim $TeV, but FIMP realization restricts it within 
$\sim 50$ GeV. The most crucial distinction however may arise from collider searches. While the WIMP realization could provide a signal excess in single or opposite sign di-lepton events associated with large missing energy, the FIMP case 
predicts stable charge track or displaced vertex signature, thanks to the production of scalar bi-doublet in the model. The decay of $\zeta_{1}^{\pm} \to \zeta_{1}^{0}$ here is restricted by the mass splitting of the order of $\delta m\sim 100$ MeV due to loop corrections. If $\delta m \lsim 0.1$ GeV, then the decay can lead to a stable charge track, while for $\delta m \lsim 0.5$ GeV, we may see displaced vertex signature. 
On the contrary, in the WIMP realization, $\zeta_{1}^{\pm}$ can easily decay to $n_{1R}$ thanks to the Yukawa coupling (which is unlikely in FIMP realization due to a heavier $n_{1R}$) and serves as an interesting 
phenomenological consequence of the model.
\section*{Acknowledgements}
M. Z. is supported by the CAS President's International Fellowship Initiative (PIFI) grant \#2019PM0110. BB and SB acknowledges DST-INSPIRE faculty grant IFA-13 PH-57. BB would also like to thank Sreemanti Chakrabarti, Rashidul Islam and Purusottam Ghosh for useful discussions. The authors would like to thank Ernest Ma for helpful discussions.

\appendix
\section{Appendix}
\subsection{Scalar States}
\label{sec:scalar}
Using the minimization conditions in Eq.~\ref{eq:sca:min}, we find the following massless Goldstone bosons:
\begin{align}
&\frac{1}{\sqrt{v_1^2 + v_2^2}} \left(-v_1 \phi^+ + v_2 \zeta_2^+\right) \label{hplus}\\
&\frac{1}{\sqrt{v_1^2 + v_2^2}} \left(-v_1 \text{Im}\left(\phi ^0_2\right) + v_2 \text{Im}\left(\zeta ^0_2\right) \right) \label{eta1}\\
&\frac{1}{\sqrt{u_2^2+2 u_3^2+v_2^2}}\left(v_2 \zeta_1^0 + u_2\chi_1  + \sqrt{2} u_3 \Delta_2 \right) \label{ksi}\\
&\frac{1}{\sqrt{u_2^2+4 u_3^2+v_1^2}} \left(v_1\text{Im}\left(\phi ^0_2\right) -u_2\text{Im}\left(\chi_2\right) +2 u_3\text{Im}\left(\Delta_3\right) \right)\label{eta2},
\end{align}
where the orthogonal state to \ref{hplus} and \ref{ksi} yields a heavy charged scalar field ($H^+$), and a physical complex scalar field ($\xi_1^0$) respectively. The orthogonal state to \ref{eta1} \& \ref{eta2} yields another physical real scalar field ($\eta^0$).\\

In the Higgs sector, the $4\times4$ mixing matrix spanning $\left( \text{Re}(\phi_2), \text{Re}(\zeta_2^0),\text{Re}(\chi_2), \text{Re}(\Delta_3) \right)$
 is given by
\begin{equation}
\left(
\begin{array}{cccc}
 2 v_1^2 \lambda _2-\frac{u_2 v_2 \mu _1}{v_1} & u_2 \mu
   _1 & 2 f_5 u_2 v_1 & 2 f_8 u_3 v_1 \\
 u_2 \mu _1 & 2 v_2^2 \left(\lambda _1+\lambda
   _3\right)-\frac{u_2 v_1 \mu _1}{v_2} & 2 f_1 u_2 v_2 & 2 \left(f_9+f_{10}\right) u_3 v_2 \\
 2 f_5 u_2 v_1 & 2 f_1 u_2 v_2 & 2 u_2^2 \lambda _4-\frac{v_1 v_2 \mu _1}{u_2} & 2 u_2\left(u_3(f_6+f_7)+\mu_{2}\right)\\
 2 f_8 u_3 v_1 & 2 \left(f_9+f_{10}\right) u_3 v_2 & 2 u_2 \left(u_3(f_6+f_7)+\mu_{2}\right) & 2 u_3^2 \left(\lambda _5+\lambda _6\right)-\frac{u_2^2 \mu _{2}}{u_3} \\
\end{array}
\right)
\end{equation}
which approximately yields a $2\times2$ block spanning $\left(\text{Re}(\phi_2), \text{Re}(\zeta_2^0)\right)$ given by:
 \begin{equation}
M_{\left\{\phi_2^R, \zeta_2^{0R}\right\}}^2 \approx \left(
\begin{array}{cc}
 2 v_1^2 \lambda _2-\frac{u_2 v_2 \mu _1}{v_1} & u_2 \mu _1 \\
 u_2 \mu _1 & -\frac{u_2 v_1 \mu _1}{v_2} \\
\end{array}
\right)
\end{equation}
The resulting mass eigenstates are given in Sec.~\ref{sec:model}. We have also used the fact that the mixing term for $\left(\text{Im}(\chi_2), \text{Im}(\Delta_3) \right)$ is negligible compared to $\Delta_3$'s mass term, and so $\Delta_3$ remains complex, with $m_{\Delta_3}^2 \approx \,\,  2u_2^2\, \mu_{2} /u_3$.

\subsection{Evolution of chemical potential}
\label{sec:chempot}

Let $d^3p~d^3x$ be the phase space element at temperature $T$ and $d^3p^{'}~d^3x^{'}$ is that at temperature $T_D$. Since the distance scales as $R$ and the momentum (of a free particle) scales as $R^{-1}$, we can write:

\bea
d^3p^{'} d^3x^{'} = \left(\frac{R}{R_D}\right)^3 d^3p \left(\frac{R_D}{R}\right)^3 d^3x = d^3p d^3x.
\eea

Since $N=n.R^3$ is fixed we have

\bea
f(p) d^3p d^3x = f(p^{'}) d^3p^{'} d^3x^{'} = f(p^{'}) d^3p d^3x \implies exp\left(\frac{E-\mu}{T}\right)=exp\left(\frac{E^{'}-\mu_D}{T_D}\right).
\eea

We consider two cases:

\begin{itemize}
 \item[(a)] Hot Relic: If particles are {\it relativistic} at the time of decoupling ($E\simeq p$), then:
 
 \bea
 exp\left(\frac{p-\mu}{T}\right) = exp\left(\frac{p^{'}-\mu_D}{T_D}\right) = exp\left(\frac{p-(R_D/R).\mu_D}{T_D.(R_D/R)}\right).
 \eea
 
 Hence, for hot relics: $T_D = \frac{R(t)}{R(t_D)}.T$ and $T_D = \frac{T_D}{T}.\mu$.
 
 \item[(b)]Cold Relic: If the particles are {\it non-relativistic} at the time of decoupling, then: $E\simeq m+\frac{p^2}{2m}$.
 
 \bea
 \begin{split}
 exp\left(\frac{E-\mu}{T}\right) &= exp\left(\frac{p^2/2m+m-\mu}{T}\right)\equiv exp\left(\frac{p^{'2}/2m+m-\mu_D}{T_D}\right)\\ & = exp\left(\frac{p^2/2m+(R_D/R)^2.(m-\mu_D)}{T_D.(R_D/R)^2}\right).
 \end{split}
 \eea
 Comparing $T_D= T(\frac{R}{R_D})^2$ with $m-\mu = (\frac{R_D}{R})^2.(m-\mu_D)$ implies $\mu(t) = m+\left(\mu_D-m\right)\frac{T}{T_D}$.
\end{itemize}

\subsection{Decay Rate for $\zeta_1 \to \zeta_2 + X$}
\label{sec:decayrate}

Using
\begin{align}
\mathcal{L} \ni & \frac{i\,g_N}{\sqrt{2}}X^+\left[\overline{\zeta_2^0}\partial_{\mu}\zeta_1^0 - \zeta_1^0\partial_{\mu}\overline{\zeta_2^0}\right] + \frac{i\,g_N}{\sqrt{2}}X^-\left[\overline{\zeta_1^0}\partial_{\mu}\zeta_2^0 - \zeta_2^0\partial_{\mu}\overline{\zeta_1^0}\right]\\
+&\frac{i\,g_N}{\sqrt{2}}X^+\left[\zeta_2^+\partial_{\mu}\zeta_1^- - \zeta_1^-\partial_{\mu}\zeta_2^+\right] + \frac{i\,g_N}{\sqrt{2}}X^-\left[\zeta_1^+\partial_{\mu}\zeta_2^- - \zeta_2^-\partial_{\mu}\overline{\zeta_1^+}\right] \\
+&\frac{g_N^2}{2}\left[\zeta_1^0\overline{\zeta_1^0}+ \zeta_2^0\overline{\zeta_2^0}+\zeta_1^-\zeta_1^++\zeta_2^-\zeta_2^+\right]\, X^+X^-,
\end{align}
the amplitude squared can be written as:
\begin{equation}
\begin{split}
\overline{\mathcal{M}|^2} =& \frac{g_N^2}{2} \left[ \frac{p_X^{\mu} p_X^{\nu}}{m_X^2} - g^{\mu\nu}\right] (p_1 + p_2)_{\mu} (p_1 + p_2)_{\nu} \\
=& \frac{g_N^2}{2} \left[ \frac{\left(m_1^2-m_2^2\right)^2}{m_X^2} + m_X^2 - 2(m_1^2+m_2^2)\right],\\
=& \frac{g_N^2}{2 m_X^2} \, \lambda(m_1^2, m_X^2, m_2^2)
\end{split}
\end{equation}
where we used
\begin{align}
E_X =& \frac{m_1}{2} \left[ 1+ \frac{m_X^2 - m_2^2}{m_1^2}\right], \\
p_f =& \frac{\sqrt{\lambda(m_1^2, m_X^2, m_2^2)}}{2 m_1},\\
\lambda(a,b,c) \equiv & (a-b-c)^2 - 4bc,
\end{align}
and so we have
\begin{align}
(p_1+p_2).p_X =&  2m_1 E_X - m_X^2 = m_1^2 - m_2^2,\\
(p_1+p_2)^2 =&  2(m_1^2 + m_2^2) - m_X^2,
\end{align}
therefore the decay rate is given by
\begin{equation}
\begin{split}
\Gamma = \left( \frac{g_N^2}{32 \pi} \right) \frac{\lambda(m_1^2, m_X^2, m_2^2)^{\frac{3}{2}}}{m_1^3 m_X^2}.
\end{split}
\end{equation}

\subsection{Decoupling time for $\zeta_1 \zeta_1^{*}\leftrightarrow$ SM}
\label{sec:decoupletime}
\noindent For the cold relic $\zeta_1$, the decoupling time $x_D$ can be determined using~\cite{Kolb:1990vq}:

\begin{equation}
x_D = \ln \left[\Lambda\right]-\frac{1}{2} \ln \left(\ln \left[\Lambda \right] \right),
\end{equation}

\noindent where $\Lambda$ is given by:

\bea
\Lambda = 0.038\, \frac{g_{\zeta _1} m_{\zeta _1}
   M_{\text{Pl}}}{\sqrt{g_{\star}}}  \sigma(\zeta_1\zeta_1 \to \rm SM) 
\eea

\noindent and we assume the decoupling occurs before the EWSB so this cross section can be written as:
\bea
\sigma(\zeta_1\zeta_1 \to \rm SM)  = \sigma(\zeta_1\zeta_1 \to \nu_L \nu_L) +  \sigma(\zeta_1\zeta_1 \to (W_3, B) \to W^+ W^-)
\eea

\noindent where we use:

\begin{align}
 \sigma(\zeta_1\zeta_1 \to \nu_L \nu_L) =&\, \frac{f_{\zeta }^2 m_{\zeta _1}^2}{8 \pi  \left(m_{\zeta
   _1}^2+m_n^2\right)^2},\\
\sigma(\zeta_1\zeta_1 \to W^+ W^-) =&\, \frac{g_2^4}{32 \pi  m_{\zeta _1}^2}
\end{align}
Since $n$ is very heavy the gauge interaction dominates the cross section.

\bea\begin{split}
f_5 u_2^2+f_8 u_3^2+f_4 v_2^2+\mu _2^2+\frac{\mu _1 u_2 v_2}{v_1}+\lambda _2 v_1^2 =0 &\\ \mu _{\zeta }^2+f_1 u_2^2+f_9 u_3^2+f_{10} u_3^2+f_4 v_1^2+\frac{\mu _1 u_2 v_1}{v_2}+\lambda _1 v_2^2+\lambda _3 v_2^2=0 &\\ \mu _{\chi}^2+f_6 u_3^2+f_7 u_3^2+f_5 v_1^2+f_1 v_2^2+\lambda _4 u_2^2+2 \mu _{23} u_3+\frac{\mu _1 v_1 v_2}{u_2}=0 &\\ \mu _{\Delta}^2+f_6 u_2^2+f_7 u_2^2+f_8 v_1^2+f_9 v_2^2+f_{10} v_2^2+\lambda _5 u_3^2+\lambda _6 u_3^2+\frac{\mu _{23} u_2^2}{u_3}=0
\end{split}\eea

\begin{equation*}
\begin{split}
f_5 u_2^2+f_8 u_3^2+f_4 \left(v_2^2+v_{\zeta_1}^2\right)+\mu _2^2+\frac{\mu _1 u_2 v_2}{v_1}+\lambda _2 v_1^2=0 &\\ \mu _{\zeta }^2+f_1 u_2^2+f_9 u_3^2+f_{10} u_3^2+f_4 v_1^2+\frac{\mu _1 u_2 v_1}{v_2}+\lambda _1 v_2^2+\lambda _3 v_2^2+\lambda _1 v_{\zeta_1}^2+\lambda _3 v_{\zeta_1}^2=0 &\\ \mu _{\chi}^2+f_6 u_3^2+f_7 u_3^2+f_5 v_1^2+f_1 v_2^2+f_2 v_{\zeta_1}^2+\lambda _4 u_2^2+2 \mu _{23} u_3+\frac{\mu _1 v_1 v_2}{u_2}=0 &\\ \mu _{\Delta}^2+f_6 u_2^2+f_7 u_2^2+f_{10} \left(v_2^2-v_{\zeta_1}^2\right)+f_9 \left(v_2^2+v_{\zeta_1}^2\right)+f_8 v_1^2+\lambda _5 u_3^2+\lambda _6 u_3^2+\frac{\mu _{23} u_2^2}{u_3}=0 &\\ \mu _{\zeta }^2+f_2 u_2^2+f_9 u_3^2-f_{10} u_3^2+f_4 v_1^2+\lambda _1 v_2^2+\lambda _3 v_2^2+\lambda _1 v_{\zeta_1}^2+\lambda _3 v_{\zeta_1}^2=0
\end{split}
\end{equation*}

\bibliographystyle{utphys}
\bibliography{Bibliography}
\end{document}